\documentclass[lettersize,journal]{IEEEtran}
\usepackage{amsmath,amsfonts}
\usepackage[caption=false,font=normalsize,labelfont=sf,textfont=sf]{subfig}
\usepackage{textcomp}
\usepackage{stfloats}
\usepackage{url}
\usepackage{verbatim}
\usepackage{CJKutf8}
\usepackage{booktabs}
\usepackage{algorithm}  
\usepackage{algorithmic} 
\usepackage{enumitem}
\usepackage{makecell}
\usepackage{threeparttable}  
\usepackage{array,xcolor,colortbl,graphicx,multirow}

\def\BibTeX{{\rm B\kern-.05em{\sc i\kern-.025em b}\kern-.08em
    T\kern-.1667em\lower.7ex\hbox{E}\kern-.125emX}}
\usepackage{balance}
\begin{document}
\title{
AgentVNE: LLM-Augmented Graph Reinforcement Learning for Affinity-Aware Multi-Agent Placement in Edge Agentic AI\\
}


\author{
    Runze Zheng, Yuqing Zheng, Zhengyi Cheng, Long Luo, Haoxiang Luo, \\
    Gang Sun, \IEEEmembership{Senior Member, IEEE}, Hongfang Yu, \IEEEmembership{Senior Member, IEEE} and Dusit Niyato, \IEEEmembership{Fellow, IEEE}
    \thanks{This work was supported in part by National Natural Science Foundation of China (62394324).} 
    \thanks{R. Zheng, Y. Zheng, Z. Cheng, L. Luo, H. Luo, G. Sun, and H. Yu are with the School of Information and Communication Engineering, University of Electronic Science and Technology of China, Chengdu 611731, China (e-mail: zrz\_uestc@163.com; zyq\_uestcz@163.com; chengzywork@163.com; llong@uestc.edu.cn; lhx991115@163.com; \{gangsun, yuhf\}@uestc.edu.cn).}
    
    \thanks{D. Niyato is with the College of Computing and Data Science, Nanyang Technological University, Singapore 639798 (e-mail: dniyato@ntu.edu.sg).}

    \thanks{The corresponding authors: Long Luo and Haoxiang Luo .}
}

\maketitle

\begin{abstract}

The Internet of Agents is propelling edge computing toward agentic AI and edge general intelligence (EGI). However, deploying multi-agent service (MAS) on resource-constrained edge infrastructure presents severe challenges. MAS service workflows are driven by complex cross-node interactions, dynamic memory accumulation, and collaborative tool usage. 
Exhibiting chain-like topological dependencies and strict affinity constraints, these workflows demand real-time responsiveness that exceeds the capabilities of traditional VNE algorithms designed for static resources. To address this, we propose \textbf{AgentVNE}, a cloud-edge collaborative framework utilizing a dual-layer architecture.
First, AgentVNE employs a large language model (LLM) to identify implicit semantic constraints and generate affinity-based resource augmentation to resolve physical dependency issues. Second, it constructs a resource similarity-aware neural network, utilizing a pre-training and PPO fine-tuning strategy to precisely capture topological similarities between dynamic workflows and heterogeneous networks. 
By coupling semantic perception with topological reasoning, this mechanism effectively bridges the gap between dynamic service requirements and physical infrastructure. Simulation results demonstrate that AgentVNE reduces workflow communication latency to less than 40\% of baselines and improves the service acceptance rate by approximately 5\%--10\% under high-load scenarios. Ultimately, this work provides a foundational solution for the semantic-aware deployment of agentic AI.

\end{abstract}

\section{Introduction}
\label{sec:introduction}


\subsection{Background}
\begin{CJK*}{UTF8}{gbsn}
With the convergence of 6G networks and Internet of Things (IoT) technologies, the computing paradigm is undergoing a fundamental shift from centralized cloud intelligence to decentralized edge general intelligence (EGI) \cite{zhang2025toward}, \cite{li2024agent}, \cite{zheng2025exploring}. Traditional edge intelligence often relies on static, task-specific models, making it difficult to cope with dynamic edge environments \cite{lin2021optimizing}, \cite{taleb2025survey}, \cite{wang2025adaptive}. However, breakthrough advancements in multimodal large language models (LLMs) have catalyzed the emergence of agentic AI. Unlike traditional AI that passively responds to instructions, agentic AI possesses a closed-loop capability of Perception-Reasoning-Action, enabling it to autonomously decompose complex goals, utilize tools, and perform multi-step reasoning \cite{zhao2025agentification}, \cite{gao2025survey}. This leap in capability drives the evolution from the Internet of Everything (IoE) to the Internet of Agents (IoA) \cite{chen2025internet}, \cite{wang2025internet}.


In the vision of the IoA, billions of heterogeneous agents would coexist at the network edge, collaborating autonomously via semantic-aware communication protocols \cite{wang2025internet}. 
According to Gartner forecasts\footnote{www.gartner.com/smarterwithgartner/what-edge-computing-means-for-i
nfrastructure-and-operations-leaders}, by 2028, at least 15\% of daily work decisions will be made autonomously by AI agents. 
This proliferation necessitates a paradigm shift from the traditional passive IoT to a proactive IoA, where edge nodes are empowered to perceive, reason, plan, and act independently in real-time through continuous perception-reasoning-action loops \cite{zhang2025toward}, \cite{wang2025internet}.
Crucially, this agentification process serves as a fundamental enabler for EGI, allowing resource-constrained devices to achieve multi-task generalization and cognitive autonomy, thereby significantly reducing reliance on centralized cloud infrastructures while enhancing adaptability in dynamic environments \cite{He2025Road}, \cite{shen2024large}.

The deployment strategies of agentic AI across the cloud-edge-end continuum significantly impact service latency and quality. Although service placement in edge computing and VNE has been widely studied, existing solutions struggle to adapt to the unique workload characteristics and lifecycle dynamics of multi-agent service (MAS) \cite{wang2025adaptive}. Traditional VNE algorithms are predominantly based on static topology assumptions, presupposing that service requests possess fixed resource demands and dependency graphs \cite{zhang2024qos}, \cite{yao2018novel}, \cite{wang2025virne}. However, agent systems reason through planning modules, where execution paths and required sub-tasks are dynamically generated at runtime \cite{zhang2025evoflow}, \cite{zhao2025agentification}. 
This high uncertainty triggers real-time deformation of the service topology during the inference process. 
Consequently, traditional algorithms relying on static pre-planning become ineffective. 
In particular, they fail to adapt to the resource demand drift induced by agent self-iteration \cite{nguyen2024dynamic}.


To date, the industry lacks specialized research on embedding multi-agent workflows in cloud-edge-endpoint environments. 
Existing studies exhibit the following limitations: Lin et al.~\cite{lin2021optimizing} investigated service placement and resource allocation in mobile edge intelligence, but focused on general computation time and energy optimization, neglecting the collaboration logic between agents; 
Wang et al.~\cite{wang2025adaptive} combined ant colony algorithms or caching strategies for AI Agent placement, but often considered only single-agent capacity matching or specific content access costs, lacking modeling of holistic MAS characteristics.
Chen et al.~\cite{chen2024spaceedge} and Li et al.~\cite{li2024incentive} proposed efficient task offloading paradigms, but their task models were limited to single-node offloading, failing to meet the essential communication and collaboration needs between agents. 
Furthermore, Fan et al.~\cite{fan2025dynamic} proposed a dynamic resource allocation scheme for distributed training. 
However, due to the fundamental differences in traffic patterns between training tasks and agent inference tasks, this method cannot be directly migrated to agentic system deployment scenarios.

\end{CJK*}

\subsection{Research Motivations}
Addressing these issues is not merely an optimization of existing network slicing or resource scheduling technologies, but a key cornerstone for building the future intelligent society.
First, it is essential for guaranteeing the scalability and real-time performance of EGI. As agent services grow exponentially, edge networks face immense pressure regarding computing power and bandwidth. Efficient topology embedding strategies are critical pathways to reducing large model inference latency and improving service acceptance rates. Researching agent-oriented network embedding can promote the dynamic self-organization and semantic interconnection of heterogeneous agents on physical networks, driving the evolution from a human-centric Internet to an agent-centric infrastructure \cite{li2024agent}, \cite{wang2025internet}.
Second, it is vital for enhancing resource efficiency and privacy security.
Precisely placing inference and memory modules at the user-side edge significantly reduces the bandwidth consumption of transmitting data to the cloud. 
Moreover, this approach ensures that sensitive context data remains within the local trusted domain. 
Consequently, it satisfies the increasingly stringent privacy compliance requirements in the IoA ecosystem~\cite{luo2024bc4llm}.

However, efficiently deploying AI agents in edge intelligence systems faces three severe challenges:

\begin{itemize}
    \item \textbf{Challenge 1: Dual Heterogeneity of Virtual Nodes and Physical Infrastructure Resources}. The deployment environment for agent services is no longer a cloud data center with uniform hardware, but a complex cloud-edge network composed of mobile devices, base stations, and edge servers. Underlying physical nodes vary vastly in multi-dimensional resource ratios, including computing, storage, and communication resources. Simultaneously, in the IoA, agents' roles are highly subdivided, leading to distinct resource demand ratios \cite{chen2024persona}, \cite{chen2025internet}. This dual resource heterogeneity on both the demand and supply sides significantly increases the difficulty of resource matching \cite{wu2024ai}. Thus, it requires deployment strategies with finer granularity to maximize system throughput.

    \item \textbf{Challenge 2: High Dynamics of Topology Structure and Resource Demand}. Unlike the relatively fixed resource partitioning in traditional VNE problems, the deployment of MAS possesses high uncertainty. On the one hand, as the system operates, the agent's memory module continuously accumulates, leading to dynamic changes in resource occupancy \cite{zheng2025exploring}, \cite{xu2025serving}. More critically, the MAS may automatically generate sub-tasks or invoke external tools during inference. Then, the system will make the service topology to dynamically expand at runtime, creating additional node placement requirements. Furthermore, the frequent joining and leaving of physical nodes in edge networks requires placement algorithms to timely cope with environmental changes and make rapid, adaptive adjustments \cite{zheng2024dhbn}.
    
    \item \textbf{Challenge 3: Hard Physical Constraints and Specific Resource Affinity.} Some agent tasks often have strong dependencies on underlying hardware, a constraint rarely considered in traditional VNE. For instance, for privacy compliance, agent nodes involving sensitive data must be placed on physical nodes with Trusted Execution Environments (TEE) \cite{luo2025toward}, \cite{luo2025weighted}; similarly, video stream processing Agents must be deployed on edge nodes connected to cameras to reduce bandwidth consumption. How to fully satisfy these hard constraints while balancing global service quality and latency optimization during large-scale node mapping is an obstacle that traditional algorithms find difficult to surmount.
\end{itemize}

\begin{figure}
    \centering
    \includegraphics[width=0.98\linewidth]{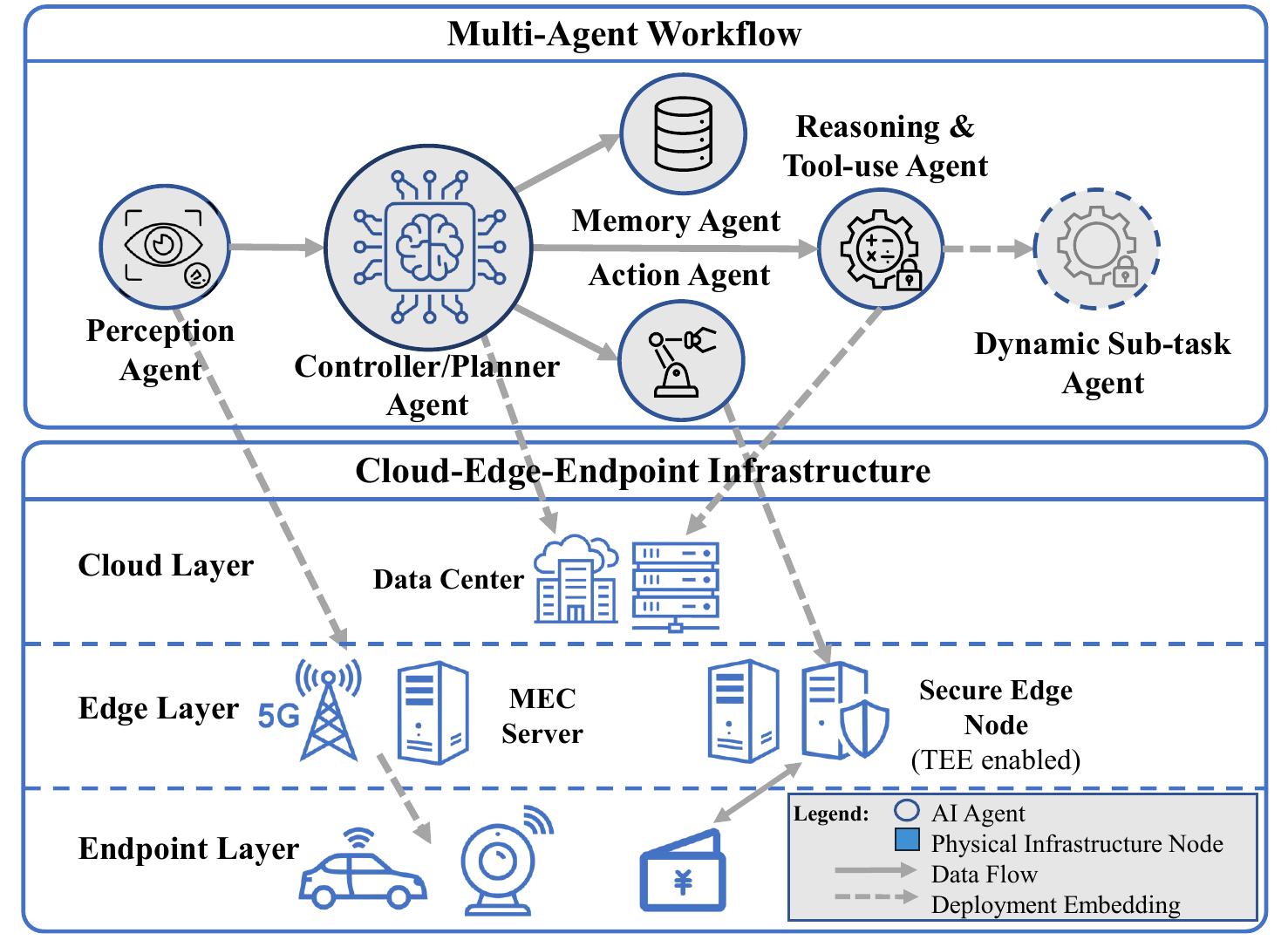}
    \caption{An illustration of the MAS embedding scenario within the Cloud-Edge-Endpoint collaborative infrastructure. It depicts the mapping of heterogeneous autonomous agents  onto physically distributed resources ranging from cloud data centers to edge devices.}
    \label{fig:placeholder}
\end{figure}
\subsection{Our Contributions}

To address the aforementioned challenges, this paper proposes an MAS embedding method oriented towards cloud-edge collaboration, \textbf{AgentVNE}. 
This method adopts a dual-layer architecture design. 
The first layer is \textbf{LLM-assisted constraint recognition and resource bias generation}, which leverages LLMs to identify task constraints and generate resource bias\footnote{Resource bias refers to an LLM-generated guidance signal that virtually augments the attributes of specific physical nodes to prioritize them for hardware affinity constraints. A formal definition is provided in Section III-B-2.} for physical nodes. 
The second layer is \textbf{deep reinforcement learning (DRL)-based topology embedding}, which utilizes deep neural networks to extract topological features and optimize placement strategies for minimizing service latency. To the best of our knowledge, this is the first work to investigate the affinity-aware service embedding problem specifically for MAS in cloud-edge-endpoint collaborative networks.
The main contributions of this paper are summarized as follows:

\begin{itemize}
    \item \textbf{An LLM-Assisted Physical Constraint Perception Mechanism:} 
    To address the strong dependency of agent systems on specific hardware, we propose an LLM-assisted judgment mechanism. 
    This mechanism automatically identifies constraint relationships between agent nodes and physical nodes and invokes a bias generator to create a resource bias for target physical nodes. 
    Specifically, this mechanism concentrates the sampling probability density around the affinity-constrained nodes and their topological neighbors. This ensures that these spatial regions are assigned higher priorities during the node selection process, thereby inducing the workflow placement to gravitate toward the affinity-constrained physical nodes.

    \item \textbf{Graph Similarity-Based Neural Network:} We design a deep neural network model fusing a column-wise neural tensor network and self-attention mechanisms. This architecture extracts features of the agentic workflow (virtual network) and substrate network through parallel graph neural network (GNN) branches, respectively. Furthermore, it utilizes a Transformer encoder to capture long-range dependencies, calculating a deep similarity matrix between nodes. This enables the model to precisely perceive the degree of matching between the physical node resource distribution and the virtual topology structure, effectively coping with the dual heterogeneity challenge.

    \item \textbf{Two-Stage Evolutionary Learning Paradigm:} 
    To overcome convergence difficulties caused by the vast solution space, we adopt a supervised pre-training and fine-tuning training strategy. 
    First, we use heuristic algorithms to generate large-scale datasets for pre-training, allowing the model to quickly capture basic topology mapping rules. 
    Subsequently, we use Proximal Policy Optimization (PPO) to fine-tune the model in a dynamic simulation environment. 
    This enables the model to adapt to the uncertainty of task dynamic arrival and expansion, achieving global performance optimization.

    \item \textbf{Comprehensive Performance Evaluation:} Simulation results demonstrate that AgentVNE significantly outperforms existing baseline algorithms in terms of communication latency, service acceptance rate, and resource utilization. 
    Specifically, the communication latency of workflows deployed via AgentVNE is reduced to less than 40\% of that yielded by GAL-VNE and metaheuristic algorithms. 
    Under high-load scenarios, the task acceptance rate improves by approximately 5\%. 
    Furthermore, the proposed framework exhibits robust scalability, maintaining rapid solving times even in large-scale network environments.
\end{itemize}


\subsection{Structure of This Work}
The remainder of this paper is organized as follows: Section II introduces the system model and problem formulation. Section III details the system architecture and design. Section IV evaluates the proposed algorithm through extensive simulation experiments. Finally, Section VI concludes the paper. Our work will be open at: https://github.com/zrzuestc/AgentVNE.
\section{System Model and Problem Formulation}
\label{sec:System Model}

\begin{table}[t]
  \centering
  \caption{Summary of Important Notations}
  \label{tab:notesummary}
  \resizebox{0.49\textwidth}{!}{
    \begin{tabular}{ll @{\hspace{2em}} ll}
      \toprule
      \textbf{Symbol}       & \textbf{Description}              & \textbf{Symbol}      & \textbf{Description}              \\
      \midrule
      $B_{l_s}^{\text{max}}$      & Link capacity                     & $B_l^{\text{req}}$         & Bandwidth demand                  \\
      $D_{l_s}^{\text{prop}}$     & Propagation delay                 & $f_{uv}^{l}$         & Link mapping var.                 \\
      $G^S$                 & Substrate graph                   & $G^V$                & Agentic workflow                  \\
      $H_s, H_v$            & Global embeddings                 & $l_v$                & Virtual link                      \\
      $N^S, E^S$            & Substrate nodes/links             & $N^V, E^V$           & Virtual nodes/links               \\
      $n_s$                 & Substrate node                    & $n_v$                & Virtual node (Agent)              \\
      $\text{NR}(u)$             & NodeRank value                    & $P_{ij}$             & Matching probability              \\
      $\mathcal{P}_{u,v}$   & Physical path                     & $r_t$                & Weighted avg. hops                \\
      $R_{n_v}$             & Resource demand                   & $U_s, U_v$           & Local embeddings                  \\
      $W$                   & NTN tensor weights                & $x_{i,u}$            & Node mapping var.                 \\
      $Z_{ji}$              & Similarity score                  & $\Omega_{n_s}$       & Resource attributes               \\
      $\Psi_{n_v}$          & Affinity constraints              & $\sigma_{i,j}$       & Data Stream                       \\
      \bottomrule
    \end{tabular}
  }
  \vspace{-0.5em}
\end{table}

This section formally defines the MAS embedding problem within a cloud-edge collaborative environment. We model the physical infrastructure as an substrate network and the multi-agent collaborative task as a virtual network. The core of the problem lies in finding a mapping function from the virtual network to the substrate network that minimizes the end-to-end latency of the MAS and optimizes resource utilization, subject to constraints on computation, storage, communication, and specific hardware affinities.

\subsection{Substrate Network Model}
The underlying physical infrastructure hosting services consists of geographically distributed heterogeneous nodes, including centralized cloud data centers, regional Mobile Edge Computing (MEC) servers, and user-side End Devices \cite{he2024efficient}. We model the substrate network as a weighted undirected graph $G^{S} = (N^{S}, E^{S})$, where $N^{S}$ represents the set of physical nodes, and $E^{S}$ represents the set of physical links.

\subsubsection{Physical Node Attributes and Resource Vectors}
Each physical node $n_s \in N^{S}$ in the set $N^{S} = \{n_1^s, n_2^s, \dots, n_{|N^S|}^s\}$ exhibits a high degree of heterogeneity. To accurately describe its capability to host LLM agents, we define a multi-dimensional attribute tuple $\mathbf{\Omega}_{n_s}$ to characterize its resource capacity,
\begin{equation}
\mathbf{\Omega}_{n_s} = \left\langle C_{n_s}^{\text{cpu}}, C_{n_s}^{\text{mem}}, C_{n_s}^{\text{disk}}, \mathcal{L}_{n_s} \right\rangle.
\end{equation}
The academic definitions of these terms are as follows:
\begin{itemize}
    \item Computational capacity $C_{n_s}^{\text{cpu}}$: Represents the general computing capability of the node, typically measured in CPU cycles (Cycles/s) or floating point operations per second (FLOPS). This determines the rate at which the node processes tasks such as data preprocessing and routine logic control.
    \item Memory capacity $C_{n_s}^{\text{mem}}$: Represents the available memory capacity (GB) of the node. Each agent must individually manage its own unique context, the length of which fluctuates dynamically during interactions. 
    \item  Storage capacity $C_{n_s}^{\text{disk}}$: Represents the persistent storage space of the node, used for storing massive knowledge bases, long-term memory logs, and container images.
    \item Location \& Affinity attributes $\mathcal{L}_{n_s}$ : This is a discrete set of attributes describing the physical location features or special hardware functionalities of the node, which is key to handling hard constraints. It is defined as
    \begin{equation}
    \mathcal{L}_{n_s} \subseteq \{ \text{Camera}, \text{TEE}, \text{LiDAR}, \dots\}.
    \end{equation}
    For example, the label $\text{TEE}$ indicates that the node supports Intel SGX or ARM TrustZone, enabling it to run privacy-sensitive agents \cite{luo2025toward}; the label $\text{Camera}$ indicates that the node is physically connected to camera sensors.
\end{itemize}

\subsubsection{Physical Link Attributes and Communication Model}
Each link $l_s = (u, v) \in E^{S}$ in the physical link set $E^{S}$ represents a communication channel between two physical nodes. Each link $l_s$ is defined by an attribute vector $\mathbf{\Phi}_{l_s}$,
\begin{equation}
\mathbf{\Phi}_{l_s} = \left\langle B_{l_s}^{\text{max}}, D_{l_s}^{\text{prop}} \right\rangle.
\end{equation}
\begin{itemize}
    \item Bandwidth capacity $B_{l_s}^{\text{max}}$ : This parameter represents the maximum bandwidth capacity of the link (Mbps). In cloud-edge collaborative environments, core network links usually possess extremely high $B^{\text{max}}$, whereas wireless links from the edge to the endpoint are relatively constrained and precious.
    \item Propagation delay $D_{l_s}^{\text{prop}}$:  The fixed propagation delay caused by signal transmission through the physical medium, primarily determined by physical distance.
\end{itemize}

In the substrate network, the communication path $\mathcal{P}_{u,v}$ between any two non-directly connected nodes $u$ and $v$ consists of a series of connected physical links. We assume the substrate network runs OSPF; therefore, the equivalent bandwidth and latency for $l_s = (u, v) \in E^{S}$ are pre-calculated for simulation purposes.

\subsection{Multi-Agent Service Network Model}
Upper-layer applications are modeled as MAS. Unlike traditional microservice architectures, nodes in the MAS are Agents with autonomous reasoning capabilities, and their interactions constitute a directed workflow \cite{qian2024scaling}. We consider an MAS service request to be a request to deploy a directed graph, denoted as $G^{V} = (N^{V}, E^{V})$.

\subsubsection{Virtual Nodes (Agents) and Dynamic Resource Demands}
The set $N^{V} = \{n_1^v, n_2^v, \dots, n_{|N^V|}^v\}$ contains all agents participating in the task (e.g., planner, coder, and reviewer). Each virtual node $n_v \in N^{V}$ represents a functional unit, and its resource demand is defined as $\mathbf{R}_{n_v}$,
\begin{equation}
\mathbf{R}_{n_v} = \left\langle R_{n_v}^{\text{cpu}}, R_{n_v}^{\text{mem}}, R_{n_v}^{\text{disk}}, \Psi_{n_v} \right\rangle.
\end{equation}
In this model, we must emphasize the dynamism and specificity of LLM agent resource demands:
\begin{itemize}
    \item $R_{n_v}^{\text{mem}}$: The memory usage of an agent. Instead, it fluctuates slightly as the task proceeds \cite{wang2025adaptive}.
    \item $\Psi_{n_v}$: This is the set of hard affinity constraints of the agent regarding the underlying nodes. If agent $n_v$ processes highly confidential data, then $\text{TEE} \in \Psi_{n_v}$ \cite{luo2025weighted}; if $n_v$ is responsible for real-time video stream analysis, then $\text{Camera} \in \Psi_{n_v}$ \cite{yang2025drivearena}. These constraints define mandatory mapping rules from virtual nodes to physical nodes.
\end{itemize}

\subsubsection{Latency Modeling}

The set $E^{V}$ represents the logical interaction relationships between agents. A directed virtual link $l_v = (i, j) \in E^{V}$ represents a communication demand from agent $i$ to agent $j$, where the content of the communication is data or a prompt processed by that Agent.

\begin{itemize}
    \item Communication latency $T_{\text{comm}}$:
Based on the assumption of equal division of underlying link bandwidth, we construct a communication latency model.

For a virtual link $l_v = (i, j)$, assuming agent $i$ and agent $j$ are mapped to physical nodes $u$ and $v$ respectively, and the physical path is $\mathcal{P}_{u,v}$, the communication latency consists of transmission delay and propagation delay, given as follows:
\begin{equation}
T_{\text{comm}}(i, j) = \sum_{e \in \mathcal{P}_{u,v}} \left( \frac{\sigma_{i,j}}{B_{e}^{\text{alloc}}} + D_{e}^{\text{prop}}\right),
\end{equation}
where $\sigma_{i,j}$ is the amount of data transmitted. $B_{e}^{\text{alloc}}$ is the actual bandwidth allocated by physical link $e$ to this virtual link.

\item Total service latency:
The total latency of the multi-agent service depends on the sum of the workflow's communication latency and the processing latency of all agents, given as follows:
\begin{equation}
T_{\text{total}} =  \sum_{n_v \in N^{V}} T_{\text{proc}}(n_v) + \sum_{l_v \in E^{V}} T_{\text{comm}}(l_v),
\end{equation}
where the processing latency $T_{\text{proc}}$ is primarily determined by the response speed of the LLM services invoked by the agents and the complexity of the tasks. As this latency depends on the external model's processing capability, its optimization falls outside the scope of this paper.
\end{itemize}

\subsection{Problem Formulation}
We formalize an optimization of service placement decision as a Mixed-Integer Non-Linear Programming (MINLP) problem. The objective is to find a mapping scheme $\mathcal{M}$ to embed the virtual network $G^V$ into the substrate network $G^S$.

The mapping $\mathcal{M}$ consists of two parts, including node mapping $\mathcal{M}_N: N^V \rightarrow N^S$, and link mapping $\mathcal{M}_L: E^V \rightarrow \mathcal{P}(G^S)$.

We define the binary decision variable $x_{i, u}$ to represent the node mapping relationship,
\begin{equation}
x_{i, u} = \begin{cases} 1, & \text{if node } i \in N^V \text{ is mapped to } u \in N^S, \\ 0, & \text{otherwise}. \end{cases}
\end{equation}
We define the variable $f_{uv}^{l}$ to represent the virtual link mapping relationship,
\begin{equation}
f_{uv}^{l} = \begin{cases} 1, & \text{if } l \in E^V \text{ traverses } (u,v) \in E^S, \\ 0, & \text{otherwise}. \end{cases}
\end{equation}

The specific constraints are as follows:
\subsubsection{Multi-dimensional Resource Capacity Constraint}
For any physical node $u$, the sum of the resource demands of all agents that it hosts must not exceed its physical capacity. 
\begin{equation}
\begin{split}
    \sum_{i \in N^V} x_{i, u} \cdot R_{i}^{k} \leq C_{u}^{k}, \quad \forall u \in N^S, \\
    \forall k \in \{\text{cpu}, \text{mem}, \text{disk}, \text{acc}\}.
\end{split}
\end{equation}

\subsubsection{Uniqueness Constraint}
Each virtual agent must be mapped to exactly one physical node,
\begin{equation}
\sum_{u \in N^S} x_{i, u} = 1, \quad \forall i \in N^V.
\end{equation}

\subsubsection{Hard Affinity and Hardware Dependency Constraints}
This is a key distinction between AgentVNE and traditional VNE. If agent $i$ has a specific set of hardware requirements $\Psi_{i}$, it can be placed only on physical nodes that satisfy this constraint,
\begin{equation}
x_{iu} \cdot \mathbb{I}\left(\alpha \notin \mathcal{L}_u\right) = 0, \quad \forall i \in N^V, \; \forall \alpha \in \Psi_i, \; \forall u \in N^S
\label{eq:affinity_constraint}
\end{equation}

\noindent where $\mathbb{I}(\cdot)$ is the indicator function. Eq.\eqref{eq:affinity_constraint} expresses that if the attribute set $\mathcal{L}_{u}$ of physical node $u$ does not contain the attribute $\alpha$ required by the agent (e.g., $\text{TEE}$), then $x_{i,u}$ must be 0.

\begin{figure}
    \centering
    \includegraphics[width=1.0\linewidth]{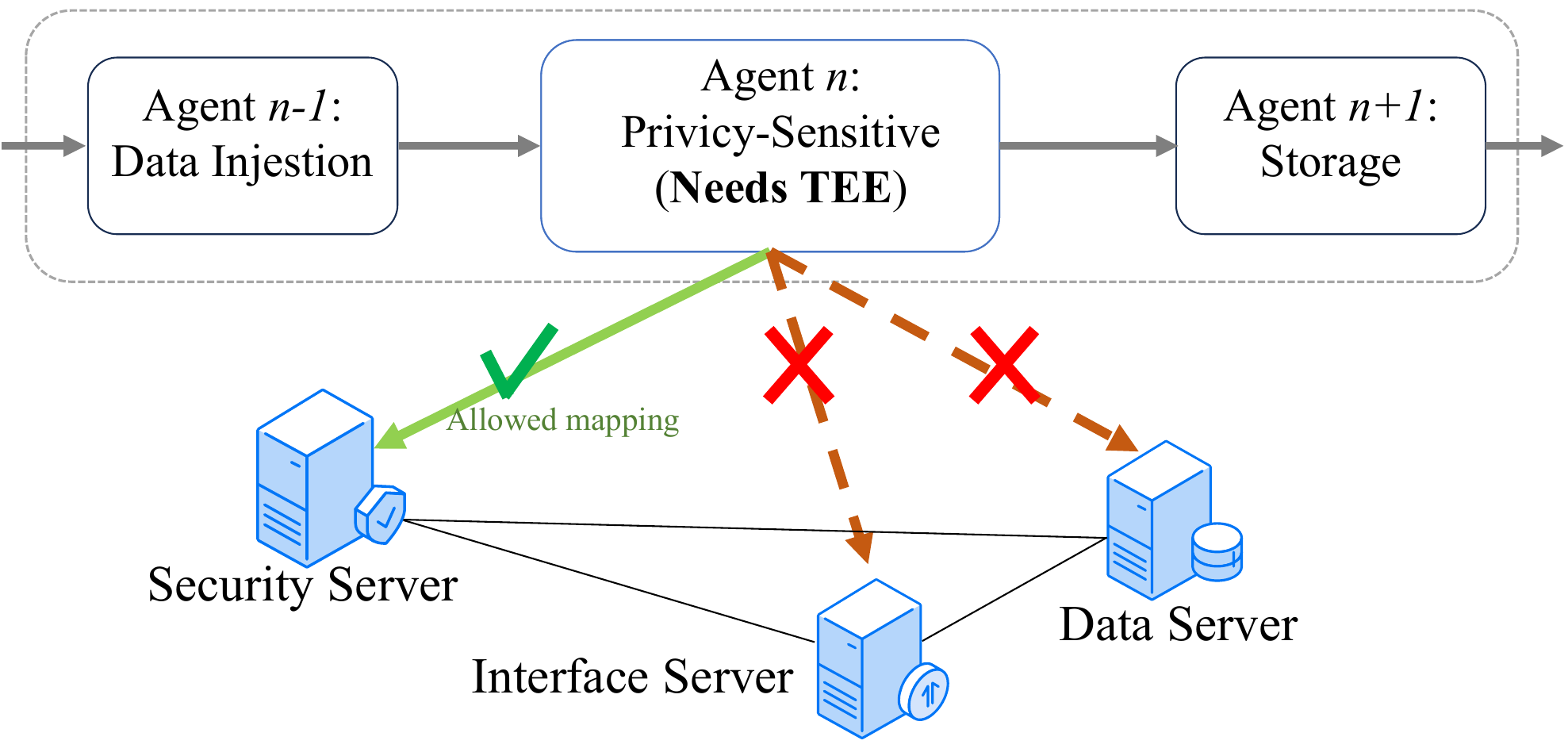}
    \caption{Schematic representation of hard physical constraints and hardware affinity in agentic workflows. Specifically, a privacy-sensitive agent requires exclusive mapping to a physical node equipped with a TEE, rendering other mappings invalid.}
    \label{fig:affinity}
\end{figure}

\subsubsection{Link Bandwidth Constraint and Bandwidth Sharing Mechanism}
Unlike bandwidth slicing in traditional VNE, data flows between agents exhibit burstiness. In the generation phase, data is transmitted as a stream; during the reasoning/thinking phase, the link remains idle. In fact, in realistic cloud-edge networks, it is rare for a service to permanently occupy the bandwidth of a link. In wireless channels at the edge, independent channels are assigned to each request via multiple access methods like OFDMA or TDMA; in the core network, bandwidth competition is managed via flow control and congestion control \cite{anitha2024comprehensive}.

When multiple virtual links are mapped onto the same physical link $e \in E^S$, they share the bandwidth $B_{e}^{\text{max}}$ of that physical link.
Let $\mathcal{F}_e = \{l \in E^V | f_{e}^{l} = 1\}$ be the set of virtual links passing through physical link $e$. Under the equal sharing model, the actual bandwidth $B_{e}^{\text{alloc}}(l)$ obtained by each virtual link is
\begin{equation}
B_{e}^{\text{alloc}}(l) = \min \left( B_{l}^{\text{req}}, \frac{B_{e}^{\text{max}}}{|\mathcal{F}_e|} \right).
\end{equation}
To ensure transmission feasibility, the physical link must be able to carry the minimum transmission requirements of all flows(or result in increased latency during congestion,
\begin{equation}
\sum_{l \in E^V} f_{e}^{l} \cdot B_{e}^{\text{alloc}}(l) \leq B_{e}^{\text{max}}, \quad \forall e \in E^S.
\end{equation}

Then, our goal is to minimize a weighted cost function, which primarily includes the MAS communication latency and the degree of resource fragmentation, as follows:
\begin{equation}
\label{formu:objection}
\begin{split}
\text{Minimize } \mathcal{J} &= \alpha \cdot \underbrace{\frac{1}{|E^V|} \sum_{(i,j) \in E^V} \text{Hops}(\mathcal{M}_N(i), \mathcal{M}_N(j))}_{\text{Communication Cost}} \\
&\quad + \beta \cdot \underbrace{\sum_{u \in N^S} \text{LB}(u)}_{\text{Load Balancing Cost}},
\end{split}
\end{equation}
where communication cost $\text{Hops}(u, v)$ represents the shortest path hop count between physical nodes $u$ and $v$. When $u=v$ (i.e., two associated agents are deployed on the same physical node), the hop count is 0, which is the ideal case as it eliminates network latency.
The load balancing cost $\text{LB}(u)$ penalizes the overload usage of nodes, encouraging balanced resource utilization to prevent the emergence of hotspots that could hinder the placement of subsequent requests.

\subsection{Problem Characteristics}
Based on the mathematical definitions above, the service embedding problem proposed by AgentVNE exhibits the following significant characteristics:

\subsubsection{NP-Hard}
This problem is reducible to the VNE problem. It is known that the VNE problem encompasses Subgraph Isomorphism \cite{cordella2004sub} and Multi-Commodity Flow \cite{szeto2003multi} problems, belonging to the NP-hard class. Furthermore, the node mapping phase involves a variant of the Quadratic Assignment Problem. The reason is that the cost of mapping agent $i$ depends on the mapping location of its neighbor agent $j$. Specifically, the distance term $\text{Hops}(\mathcal{M}(i), \mathcal{M}(j))$ in the objective function. As the number of agents $|N^V|$ and physical nodes $|N^S|$ increases, the solution space grows exponentially ($O(|N^S|^{|N^V|})$), making exact solutions infeasible within polynomial time.

\subsubsection{Sparse Feasibility Region and Hardware Affinity}
Due to the introduction of hard affinity constraints (Eq.\ref{eq:affinity_constraint}), the space of feasible solutions becomes extremely discontinuous and sparse. Specific physical nodes exert a strong attraction on specific types of virtual nodes. This effect greatly limits the degrees of freedom for search algorithms. Traditional heuristic algorithms struggle to capture such constraints, potentially leading to the premature exhaustion of resources on constrained nodes. Thus, it will increase the resource fragmentation of the underlying network and decrease service capacity \cite{wu2024ai}.

\subsubsection{Dynamic Topology Evolution}
Unlike the relatively static topology structure of traditional network services, MASs possess high dynamism. The Chain-of-Thought (CoT) \cite{wei2022chain} of agents may dynamically generate new sub-tasks during inference, causing the node set $N^V$ and edge set $E^V$ of $G^V$ to change at runtime. Therefore, the problem formulation is no longer a one-time static optimization but potentially a time-varying graph embedding \cite{qian2024scaling}. Then it will require algorithms to possess capabilities for incremental computation and rapid re-mapping.

\subsubsection{Resource Coupling}
Computational resources and network resources are tightly coupled in the optimization objective. To reduce communication latency, namely minimize hops, algorithms tend to centrally place all agents on a single powerful cloud node. However, this leads to instantaneous overloading of that node's compute and memory resources (violating resources constraint). Conversely, distributed placement for load balancing purposes sharply increases communication latency. The MINLP formulation mathematically quantifies this trade-off through weights $\alpha$ and $\beta$ in the objective function, as well as strict capacity constraints.
\section{System Architecture and Design Details}
\label{sec:proposed_approach}

This section elaborates on the system architecture and core design of AgentVNE. As shown in Fig. \ref{framework}, the system consists of two layers: (1) an LLM-based virtual-physical node correlation capture and bias generation module, and (2) a deep neural network-based decision module designed to encode heterogeneous graph features. The system generates specific node mapping schemes, and continuously optimizes policies driven by online learning.

\subsection{System Architecture and Workflow}
\label{sec:system_arch}

\begin{figure*}[t]
    \centerline{\includegraphics[width=0.84\textwidth]{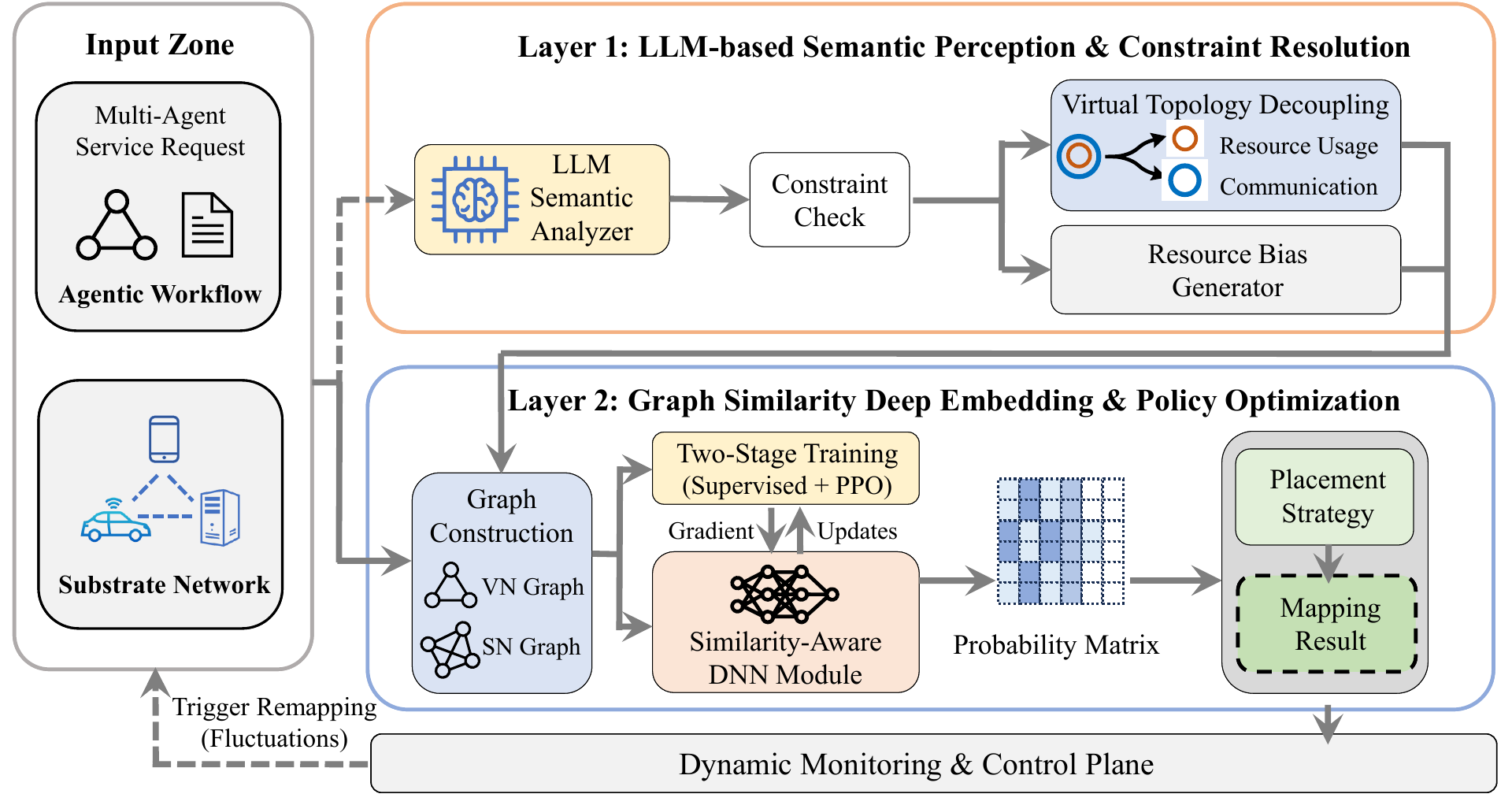}}
    \caption{The overall framework of the proposed AgentVNE. The architecture consists of two key layers. (1) An LLM-based semantic perception layer for constraint resolution and resource bias generation. (2) A graph similarity-based deep embedding layer for adaptive placement optimization.}
    \label{framework}
\end{figure*}

As illustrated in Fig. \ref{framework}, AgentVNE establishes a dual-layer framework for the adaptive embedding of self-evolving MASs. The architecture operates through a dual-layer coupling mechanism aimed at bridging the gap between semantic service requirements and heterogeneous physical infrastructure.

\subsubsection{Layer 1: Semantic Perception and Resource Bias Injection}
The first layer utilizes an LLM as a semantic parser to address the dual heterogeneity challenge. Upon receiving a MAS placement request, this layer analyzes the semantic context to identify implicit hard constraints. Instead of rigid rule matching, we introduce a topology decoupling and resource bias mechanism. This mechanism transforms semantic constraints into graph-level features by decoupling restricted agents into anchor nodes and resource nodes, while simultaneously injecting a resource bias into the substrate network. This preprocessing effectively guides the subsequent embedding probability density toward physical nodes that satisfy hard constraints.

\subsubsection{Layer 2: Deep Geometric Embedding with Evolutionary Optimization}
The second layer serves as the core decision engine, employing a similarity-aware graph neural network. Taking the preprocessed virtual network and substrate network graphs as input, this module quantifies the high-order topological similarity between the agent workflow and physical resources. To address the NP-hard nature of the embedding problem, we adopt a two-stage training paradigm consisting of supervised pre-training and PPO fine-tuning. This enables the model to rapidly capture basic topology mapping rules through supervised learning, and subsequently evolve its policy via PPO to maximize long-term returns in dynamic environments.

\subsubsection{Dynamic Closed-Loop Control}
Finally, the system integrates a dynamic monitoring plane. When topology evolution or infrastructure fluctuations are detected, the control loop triggers a re-reasoning process. For instance, this detection is triggered when new virtual nodes are generated during the execution of the multi-agent service. The updated graph state is fed back into the neural network to generate adaptive migration or expansion strategies, ensuring service continuity.

\subsection{LLM-based Semantic Perception and Constraint Resolution}
The first layer is the semantic perception and constraint resolution layer based on LLM. 
It is primarily responsible for the semantic matching of virtual nodes with physical resources and the handling of hard constraints. When a multi-agent service request arrives, the LLM first performs semantic analysis on the service type, e.g., real-time video surveillance stream, to identify whether strong resource affinity or location dependency constraints exist between virtual network and substrate network. For scenarios with hard physical constraints, such as a specific computational task that must be executed on an edge node equipped with a camera interface, the system introduces a topology reconstruction and resource injection mechanism:

\subsubsection{Virtual Topology Decoupling} The constrained virtual node is decoupled into a resource node, which executes computational tasks, and a communication anchor, which has zero resource demand. This anchor is forcibly mapped to the target physical node to anchor the spatial location of the logical topology.

\subsubsection{Resource Bias Injection} The resource bias generator is used to synthesize an affinity-based augmentation vector for the target physical node. These augmented attributes are then integrated into the feature matrix and fed into the neural network. This bias resource is then diffused around the constrained node through resource iterative computation or Graph Convolutional Networks (GCN). Therefore, this significantly improves the selection priority of the constrained node and its neighboring nodes in subsequent random sampling processes.

\subsection{Similarity-Aware Feature Extraction Network}

\begin{figure*}[t]
\vspace{-0.2em}
\centerline{\includegraphics[width=1.0\textwidth]{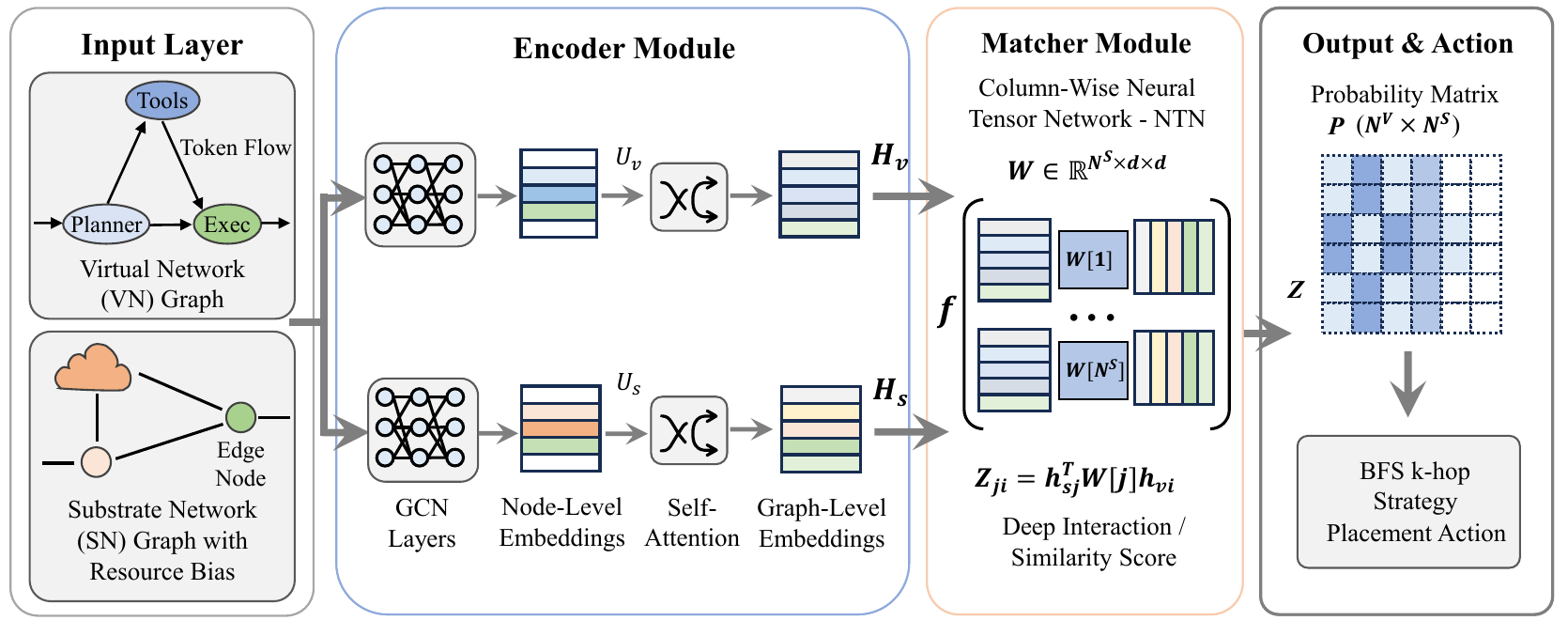}}
\caption{Detailed network structure of the similarity-aware feature extraction module. The architecture comprises dual-stream graph encoders for feature learning, a Column-Wise ntn for deep similarity interaction, and a heuristic placement strategy based on the output probability matrix.}
\label{DNN module}
\vspace{0.1em}
\end{figure*}

In traditional VNE studies, most existing algorithms employ heuristic or metaheuristic strategies \cite{geng2023gal}. They tend to prioritize mapping virtual nodes with large resource demands to physical nodes with the largest remaining capacities. This capacity-driven strategy is relatively effective in low-dimensional resource scenarios. However, it exhibits significant limitations when facing the high-dimensional heterogeneous resource demands of agent systems. It ignores the shape alignment between the virtual node demand vector and the physical node supply vector in multi-dimensional space. This oversight easily leads to the formation of unusable, fragmented resources on physical nodes, resulting in severe resource fragmentation, which in turn reduces the long-term service acceptance capability of the network \cite{geng2023gal}.

Inspired by SimGNN \cite{bai2019simgnn},
 we propose that an ideal node mapping should not only satisfy resource capacity constraints but also consider the structural and resource coherence between virtual nodes and physical nodes with respect to their multi-dimensional resource profiles and topological characteristics. To this end, we design a deep neural network with both topological awareness and resource awareness. This network captures the deep-seated structural similarity between the virtual network graph $G^V$ and the substrate physical graph $G^S$, guiding node placement toward regions with more similar resource dimension profiles and topological relationship similarity, thereby optimizing resource utilization efficiency.

The specific architecture of the network is shown in Fig. \ref{DNN module}. It takes two graphs as input and outputs an association probability matrix between node pairs.

\subsubsection{Dual-Stream Graph Encoding}
To extract high-order features from heterogeneous graphs, we design a dual-stream parallel encoding architecture. First, GCN \cite{kipf2016semi} is employed to process $G^V$ and $G^S$ separately to aggregate local neighborhood information. For a physical node $n_s$, its initial feature vector $x_{n_s}$ encapsulates the normalized residual capacities across diverse resource dimensions (e.g., CPU, memory, and bandwidth). After $L$ layers of GCN aggregation, the node obtains the locally perceived feature $U_s$,
\begin{equation}
    U_s^{(l+1)} = \sigma (\tilde{D}^{-\frac{1}{2}} \tilde{A} \tilde{D}^{-\frac{1}{2}} U_s^{(l)} W^{(l)}).
\end{equation}
Similarly, the feature representation $U_v$ for agent network nodes is obtained. 

Although GCNs excel at capturing local structures, they struggle to perceive topological dependencies across the entire graph \cite{li2018deeper}.
Therefore, we further input $U_v$ and $U_s$ into a Transformer encoder layer. Utilizing Multi-Head Self-Attention (MHSA) and Feed-Forward Networks (FFN), the model dynamically weights the correlations between different nodes, identifying global key nodes such as communication bottlenecks or resource hubs. For the virtual network feature $U_v$, its enhanced global context feature $H_v$ after passing through the Transformer encoder \cite{vaswani2017attention} is calculated follows:
\begin{equation}
    \begin{aligned}
        U'_v &= \text{LayerNorm}(U_v + \text{MHSA}(U_v)), \\
        H_v &= \text{LayerNorm}(U'_v + \text{FFN}(U'_v)).
    \end{aligned}
\end{equation}
The generation process for the substrate network feature $H_s$ is identical to that of $H_v$, with the input replaced by $U_s$. Finally, $H_v$ and $H_s$ serve as high-dimensional embedding vectors rich in global topological semantics, which are input into the subsequent neural tensor network (NTN).

\subsubsection{Column-Wise Neural Tensor Network}
After obtaining high-dimensional embeddings for the nodes, the core task is to compute the matching score between agent node $i$ and physical node $j$. 
Thus, AgentVNE introduces a column-wise NTN as the core matching module. This module performs slice-level deep interaction on features through a learnable 3D tensor $W \in \mathbb{R}^{|N^S| \times d \times d}$ as follows:
\begin{equation}
    Z_{ji} = h_j^T W_j h_i + b,
\end{equation}
where $h_j$ and $h_i$ are the enhanced feature vectors of the physical node and virtual node, respectively, and $W_j$ is the slice weight matrix in the tensor corresponding to physical node $j$. This design allows the model to learn unique resource matching patterns for each physical node \cite{bai2019simgnn}, thereby generating a high-precision raw similarity matrix $Z$.

Finally, the output layer first uses the sigmoid activation function to map the arbitrary real values of the raw similarity $Z$ to the non-linear interval $(0, 1)$ to obtain matrix $S$ as follows:
\begin{equation}
    S_{ij} = \sigma(Z_{ij}) = \frac{1}{1 + e^{-Z_{ij}}}.
\end{equation}
To eliminate dimensional differences between samples and generate a smooth probability distribution, we perform Row-wise L2 Normalization on $S$. The final matching probability $P_{ij}$ for mapping the $i$-th virtual node to the $j$-th physical node is defined as:
\begin{equation}
    P_{ij} = \frac{S_{ij}}{\| S_i \|_2} = \frac{S_{ij}}{\sqrt{\sum_{k=1}^{|N^S|} S_{ik}^2 + \epsilon}},
\end{equation}
where $\epsilon$ is a small constant. This normalized probability matrix $P_{ij}$ is directly used to generate priority lists to guide the subsequent heuristic node mapping decisions.

\subsection{Placement Strategy}

The similarity matrix $Z$ output by the network represents only potential matching probabilities. During the training phase, we sample from the row vectors of the similarity matrix for each virtual node to generate substrate node priority sequences $L_{n_v}$. In the runtime phase, the priority sequence $L_{n_v}$ is constructed by ranking all substrate nodes in descending order of their similarity probabilities.

Subsequently, we employ a heuristic search algorithm to convert continuous probabilities into discrete actions. The specific steps are as follows:

\subsubsection{Priority List Generation}
The construction of the candidate priority list $L_{n_v}$ depends on the operational phase. 
For training, multiple samples are drawn stochastically based on $P(\cdot | n_v)$. 
For inference, $L_{n_v}$ is generated by ranking all substrate nodes in descending order according to the magnitude of their matching probabilities in $P$.
\subsubsection{Greedy-based Breadth-First Search (BFS) Expansion Strategy}
We adopt a greedy initialization where the anchor agent (highest resource demand and centrality) is directly mapped to the top-ranked physical node in $L_{anchor}$. Subsequently, the remaining agents are deployed via BFS. To greedily minimize communication latency, the strategy prioritizes searching for available resources within the immediate $k$-hop neighborhood of the preceding physical node, starting with $k=1$. The search radius is incrementally expanded ($k \leftarrow k+1$) only if no feasible solution is found, ensuring a compact topology embedding. If an expansion step fails, the system triggers a rollback mechanism, releasing occupied resources, penalizing the probability of the current path.

\subsection{Two-Stage Evolutionary Training Framework}

Given the massive solution space of combinatorial optimization problems, direct use of Reinforcement Learning (RL) suffers from extremely slow cold-start convergence \cite{geng2023gal}. We adopt a two-stage training strategy, namely supervised pre-training and PPO fine-tuning.

\subsubsection{Stage 1: Topology-Aware Pre-training}
We utilize heuristic-based algorithms to generate a large number of $\langle \text{virtual network node, noderank} \rangle$ probability matrix samples. By minimizing the mean squared error (MSE) loss function, the model rapidly learns basic topology matching rules. That is, what type of agent structure should be mapped to what type of physical structure. According to the RW-BFS \cite{cheng2011virtual} and GAL-VNE \cite{geng2023gal} for the calculation of noderank to serve as labels for pre-training our network. The noderank algorithm aims to comprehensively consider the computational resources and network topology of nodes to quantify their importance in the network. The calculation process primarily consists of three phases, including initial resource assessment, iterative resource propagation, and ranking sharpening.

Firstly, we define a resource assessment score $H(u)$ for each node $u \in V$ in the network. This score reflects the raw processing and transmission capabilities of the node, determined by the product of its CPU core count and communication bandwidth as follows:
\begin{equation}
    H(u) = \text{CPU}(u) \times \text{BW}_{\text{comm}}(u),
\end{equation}
where $\text{CPU}(u)$ represents the number of CPU cores and $\text{BW}_{\text{comm}}(u)$ represents the node's communication bandwidth. By normalizing $H(u)$ across the entire network, we obtain the initial ranking distribution $\mathbf{r}^{(0)}$ from
\begin{equation}
    r^{(0)}(u) = \frac{H(u)}{\sum_{v \in V} H(v)}.
\end{equation}

Secondly to capture the influence of topological structure, we construct a forward probability matrix $P_F$. The matrix element $P_F(u, v)$ represents the probability of resource contribution from node $u$ to neighbor $v$, which is proportional to the resource share of $v$ among all neighbors of $u$,
\begin{equation}
    P_F(u, v) = 
    \begin{cases} 
    \frac{H(v)}{\sum_{k \in \mathcal{N}(u)} H(k)}, & \text{if } v \in \mathcal{N}(u), \\
    0, & \text{otherwise}.
    \end{cases}
\end{equation}
Based on this matrix, the algorithm performs iterative aggregation. In each iteration $t$, node rankings aggregate the weighted contributions of neighbor nodes by introducing a propagation factor $\mu$, 
\begin{equation}
    \tilde{\mathbf{r}}^{(t+1)} = \mathbf{r}^{(t)} + \mu (P_F \cdot \mathbf{r}^{(t)}).
\end{equation}
After each update, the vector must be normalized by $\mathbf{r}^{(t+1)} = \tilde{\mathbf{r}}^{(t+1)} / \|\tilde{\mathbf{r}}^{(t+1)}\|_1$. This process is repeated for a specific number of iterations to fully fuse local topological features \cite{cheng2011virtual}.

Finally, to effectively identify high-priority nodes through differentiation and suppress interference from resource-deficient nodes---specifically those with insufficient capacity to host any agent---we introduce a non-linear cubic transformation. The final noderank value $NR(u)$ is calculated as follows:
\begin{equation}
    NR(u) = \frac{(r(u))^3}{\sum_{v \in V} (r(v))^3}.
\end{equation}
This sharpening step amplifies numerical differences, ensuring that nodes with high-quality aggregated resources are significantly distinguishable in the final ranking.

\subsubsection{Stage 2: Dynamic PPO Fine-tuning}
Based on the pre-trained weights, we initialize the Actor-Critic architecture. 
The agent interacts with the environment and updates the policy network parameters using the PPO algorithm based on long-term cumulative rewards. This stage enables the model to adapt to system stochasticity and learn to leverage resource bias for global optimization.

The RL reward function $r_t$ aims to minimize the average communication latency of all surviving tasks in the network while maximizing the service acceptance rate. $r_t$ is defined as follows:
\begin{equation}
    r_t = 
    \begin{cases} 
    \mathcal{P}_{fail}, & \text{if embedding fails}, \\
    \mathcal{R}_{sys}(t), & \text{success or no arrival},
    \end{cases}
\end{equation}
where $\mathcal{P}_{fail}$ is a fixed penalty term for placement failure, which occurs when the system fails to locate sufficient resources to accommodate the agentic workflow. $\mathcal{R}_{sys}(t)$ represents the system operational utility at time $t$, defined as the negative average congestion-weighted hops of all surviving workflows as follows,
\begin{equation}
    \label{eq:reward}
    \mathcal{R}_{sys}(t) = - \frac{1}{|\mathcal{W}_t|} \sum_{w \in \mathcal{W}_t} \sum_{l^v \in E_w^V} \sum_{e \in \mathcal{P}(l^v)} C_e(t),
\end{equation}
where $\mathcal{W}_t$ is the set of all surviving workflows in the network at time $t$; $E_w^V$ is the set of virtual links in workflow $w$; $\mathcal{P}(l^v)$ is the underlying physical path, namely the set of physical links, to which virtual link $l^v$ is mapped; $C_e(t)$ denotes the number of workflows sharing physical link $e$ at time $t$.

By penalizing link congestion, this reward function guides agents to plan routing paths that are both short and avoid congestion. Thus, it can optimize global end-to-end latency.

\begin{algorithm}
\caption{AgentVNE Inference Algorithm}
\begin{algorithmic}[1]
\REQUIRE Substrate Network $G^S$, Agent Workflow $G^V$, Model Parameters $\theta$
\ENSURE Mapping Scheme $\mathcal{M}$

\vspace{0.1cm}
\STATE \textbf{Phase 1: Semantic Perception (LLM)}
\STATE Generate resource bias: $B_{\text{bias}} \leftarrow \text{LLM}(G^V)$
\STATE Update substrate attributes: $\Omega_{n_s} \leftarrow \Omega_{n_s} + B_{\text{bias}}$

\vspace{0.1cm}
\STATE \textbf{Phase 2: Deep Embedding \& Matching}
\FOR{$k \in \{S, V\}$}
    \STATE $U_k \leftarrow \text{GCN}(X_k, \tilde{A}_k)$; \quad $H_k \leftarrow \text{Transformer}(U_k)$
\ENDFOR
\STATE Compute interaction tensor $Z$: 
\STATE \quad $Z_{ji} \leftarrow h_j^T W_j h_i + b, \quad \forall j \in N^S, i \in N^V$
\STATE Generate probability matrix $P$:
\STATE \quad $P_{ji} \leftarrow \frac{\sigma(Z_{ji})}{\sqrt{\sum_{k} \sigma(Z_{ki})^2 + \epsilon}}$ 

\vspace{0.1cm}


\STATE \textbf{Phase 3: Greedy-BFS Placement}
\STATE Map hard constraints; 
\STATE Map anchor agent (max resource) to top-ranked node.
\FOR{each unmapped agent $v$ in BFS order}
    \STATE $u_{\text{pre}} \leftarrow \mathcal{M}(\text{Predecessor of } v)$; 
    \STATE $L_v \leftarrow \text{argsort}_{\downarrow}(P_{v, :})$.
    \STATE Search feasible $u^* \in L_v$ starting from $u_{\text{pre}}$ and incrementally expanding $k$.
    \STATE $\mathcal{M}(v) \leftarrow u^*$.
\ENDFOR

\vspace{0.1cm}
\STATE Map virtual links $\mathcal{M}_L$ using Shortest Path on $G^S$.
\RETURN $\mathcal{M}$
\end{algorithmic}
\end{algorithm}



The complete process of AgentVNE inference can be shown as Algorithm 1. The computational complexity of AgentVNE is primarily dominated by the neural network inference, specifically the self-attention and tensor matching mechanisms, which scale quadratically with the substrate network size as $O(|N^S|^2)$. With matrix overheads masked by GPU parallelism, the runtime is dominated by CPU-side logic, ensuring robust scalability.

\section{Performance Evaluation}\label{sec:performance_evaluation}
\label{sec:proposed_approach}


In this section, we evaluate the performance of AgentVNE through extensive simulations. We compare our framework against other relevant methods to analyze its efficacy in reducing latency, improving acceptance rates, and maintaining scalability.

\subsection{Experimental Setup and Baselines}

To evaluate the performance of AgentVNE, we established a cloud-edge-endpoint collaborative simulation environment. The substrate network is modeled as a weighted undirected graph consisting of heterogeneous nodes, including centralized cloud data centers, regional MEC servers, and user-side end devices. The substrate network is generated by Waxman topology model \cite{waxman2002routing} with $\alpha = 0.5, \beta = 0.2$. To reflect the distinct characteristics of the IoA, approximately $10\%$ of the substrate nodes are configured with specific hardware attributes to enforce hard affinity constraints. We utilize a locally deployed Qwen3-30B as the core for resource bias generation in the first stage.

All simulations were implemented in Python 3.12 and conducted on a high-performance server equipped with an Intel Xeon\textsuperscript{\textregistered} Silver 4310 CPU, an NVIDIA\textsuperscript{\textregistered} Tesla\textsuperscript{\textregistered} T4 GPU, and 64 GB of RAM.




\begin{table}[tbp]
\centering
\caption{Simulation Environment and Hardware Specifications}
\label{tab:sim_settings}
\renewcommand{\arraystretch}{1.2}
\begin{tabular}{ll}
    \toprule
    \textbf{Parameter} & \textbf{Value/Description} \\ 
    \midrule
    \multicolumn{2}{l}{\textit{\textbf{Substrate Network}}} \\
    Topology Structure              & Waxman topo $\alpha = 0.5, \beta = 0.2$\\
    Network Size                    & 10 $\sim$ 600 nodes\\
    Node Affinity Ratio             & $\sim$10\% Hardware affinity \\
    \midrule
    \multicolumn{2}{l}{\textit{\textbf{Agentic Workflow (VN)}}} \\
    Network Size                    & 5 $\sim$ 7 nodes\\
    Arrival Rate ($\lambda$)        & Poisson(0.05$\sim$0.5)   \\
    Mean Lifetime ($\mu$)           & Exp ($15 \sim 40$)   \\
    \bottomrule
\end{tabular}
\end{table}

Regarding service requests, we synthesize four distinct types of MAS workflows based on real-world frameworks such as LangChain \cite{chase2022langchain} to calibrate the actual resource consumption profiles. As detailed in Table \ref{table:workflow_scheme}, the test workflows encompass two primary categories: Text search and image processing. These workflows feature a node scale ranging from 5 to 7 agents and possess varying affinity requirements. In the simulation, agent task arrivals follow a Poisson process, while their service lifetimes obey an exponential distribution, simulating a dynamic execution environment \cite{baccelli2013elements}. The four workflow types follow a uniform distribution within the incoming task stream. The specific configurations of these workflows will be provided in the open-source repository.

To demonstrate the effectiveness of AgentVNE, we compare it against the following baseline algorithms:
\begin{itemize}
    \item \textbf{Greedy} \cite{gong2014toward}
    : A magnitude-centric strategy that prioritizes physical nodes with the maximum residual resources for placement.
    \item \textbf{GA (Genetic Algorithm)} \cite{dab2013vnr}: Represents traditional meta-heuristic algorithms that utilize population evolution and iterative search mechanisms to identify near-optimal mapping schemes.
    \item \textbf{GAL-VNE} \cite{geng2023gal}: A learning-based baseline employing a pre-training paradigm to calculate node scores, subsequently performing node mapping in an approximate greedy manner based on these rankings.
\end{itemize}

\begin{table}[t]
    \centering
    \begin{threeparttable}
    \caption{Workflow Type}
    \label{table:workflow_scheme} 
    \begin{tabular}{cccc} 
        \toprule 
        \textbf{Workflow ID} & \textbf{Node Number} & \textbf{Task Type\tnote{1}} & \textbf{Affinity Nodes}\\
        \midrule
        workflow\_1 & 6 & Text search & No \\
        workflow\_2 & 7 & Text search & Yes \\
        workflow\_3 & 5 & Image processing & No \\
        workflow\_4 & 6 & Image processing & Yes \\
        \bottomrule
    \end{tabular}
    \begin{tablenotes}
        \footnotesize
        \item[1] See https://github.com/zrzuestc/AgentVNE/tree/main/Workflow\_topo for details.
    \end{tablenotes}
    \end{threeparttable}
\end{table}


\subsection{Workflow Latency Analysis}

Communication latency within workflows significantly impacts the end-to-end response time of MASs. Therefore, we employ the workflow weighted average hops $r_t$ of all currently deployed tasks in the network as the core metric to measure the post-deployment communication overhead.


\begin{figure*}[htbp]
    \centering
    \subfloat[Long-term Weighted Average Hops\label{fig:vn}]{
        \includegraphics[width=0.31\textwidth]{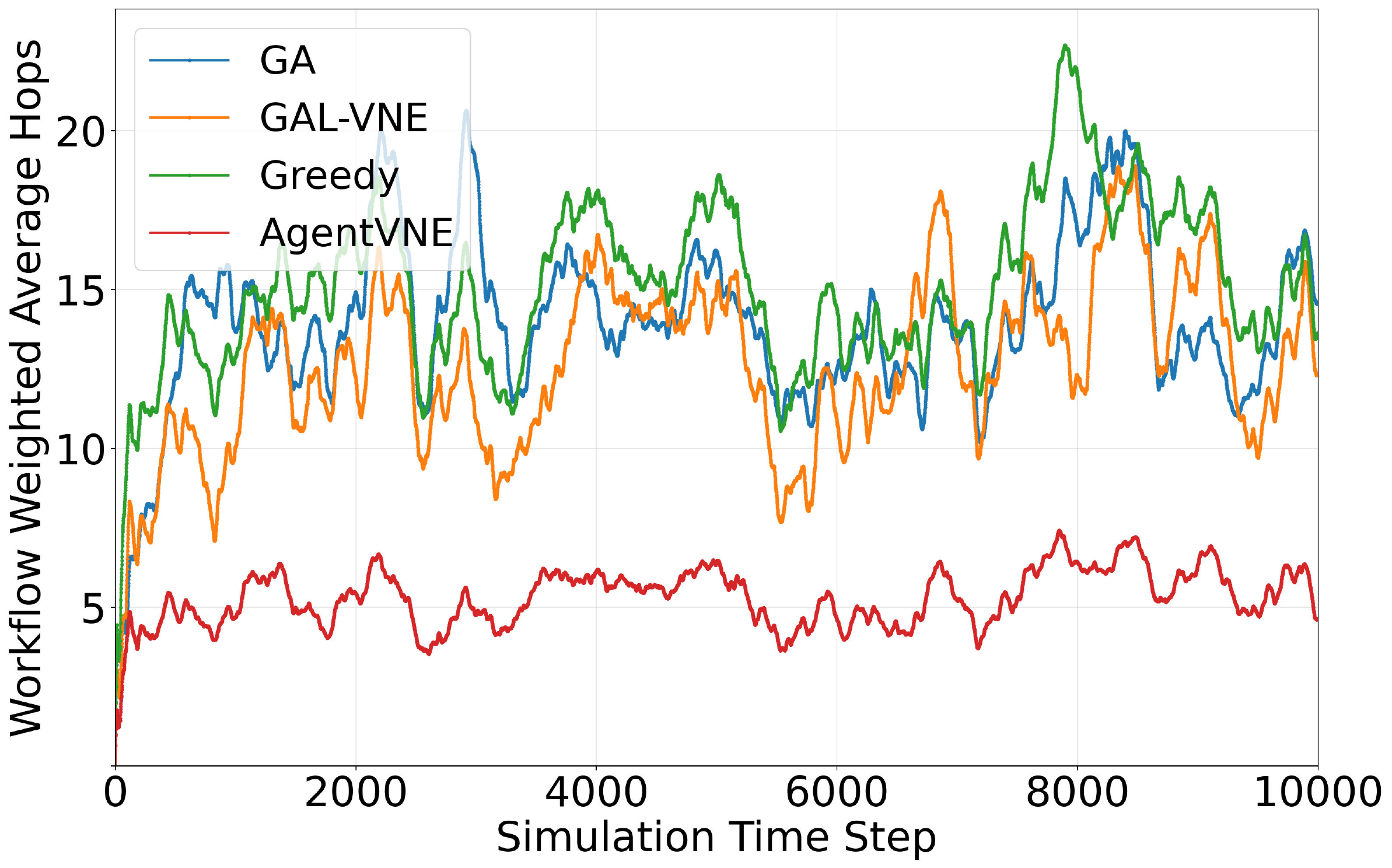}
        \label{fig:hops_time}
    }
    \subfloat[Arrival Rate\label{fig:pretrain_prob}]{
        \includegraphics[width=0.30\textwidth]{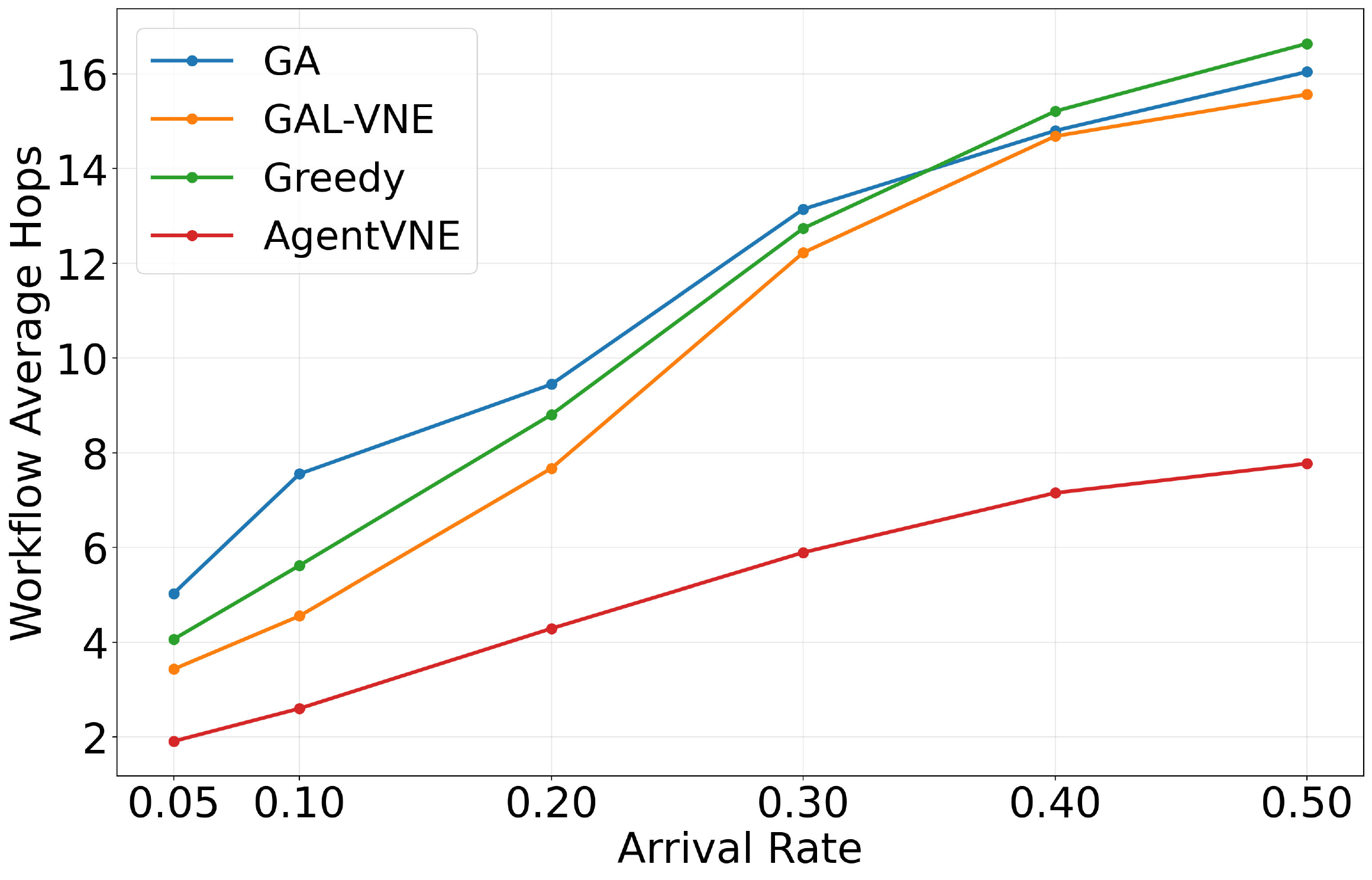}
        \label{fig:hops_time_1}
    }
    \subfloat[Workflow Mean Lifetime\label{fig:finetuned_prob}]{
        \includegraphics[width=0.31\textwidth]{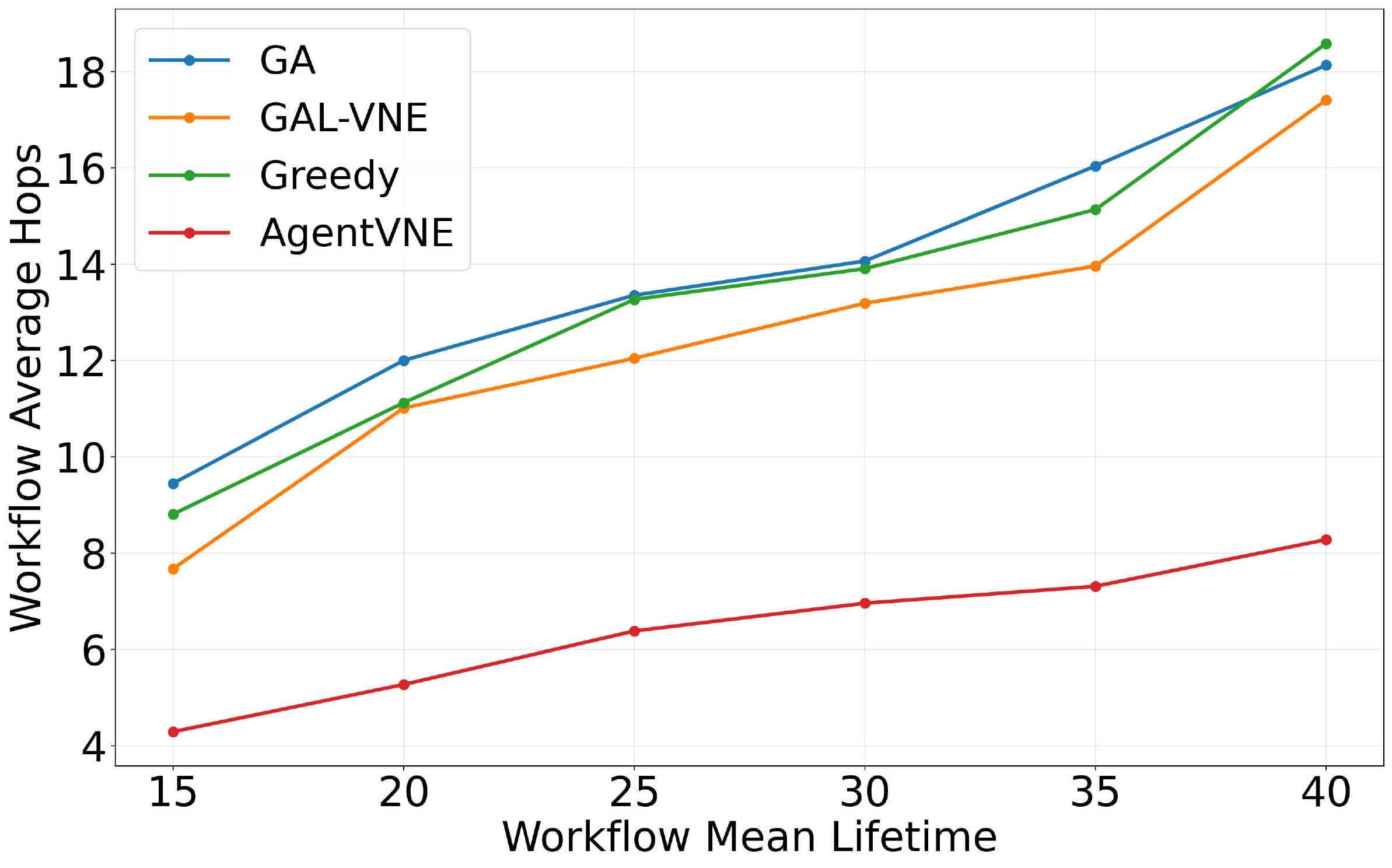}
        \label{fig:hops_time_2}
    }
    \caption{Multi-agent workflow communication latency comparison, namely weighted average hop count performance in different scenarios.}
    \label{fig:r_t}
    \vspace{-0.5em}  
\end{figure*}





\subsubsection{Temporal Stability and Topology Aggregation}
We first evaluate the stability of the algorithm over the temporal dimension. Setting the task arrival rate to 0.2 and the mean lifetime to 30 time units, the variation of $r_t$ over 10,000 simulation steps is shown in Fig. \ref{fig:hops_time}. The results indicate that AgentVNE consistently maintains the lowest average hops, exhibiting superior temporal stability.

AgentVNE outperforms baselines in two dimensions. First, it reduces the average communication hops by approximately $60\%$ compared to algorithms like GAL-VNE and GA. More crucially, the variance of its latency fluctuation is restricted to merely $20\%$ or less of the baseline algorithms. This minimal jitter demonstrates that AgentVNE provides not only lower latency but also significantly more stable and predictable communication performance.

Furthermore, this significant improvement is attributed to the structural alignment capability of our dual-stream graph encoding architecture. Traditional baseline strategies often employ a capacity-centric greedy logic. This logic tends to scatter nodes across physical nodes with the most residual resources, thereby disrupting workflow connectivity. In contrast, by integrating GNN and Transformer modules, AgentVNE captures the high-order topological isomorphism between the virtual workflow and the substrate network. This enables the algorithm to proactively preserve the locality of virtual links during the mapping process. It can ensure that logically connected agents are deployed onto physically proximal neighborhoods for compact topological embedding.


\subsubsection{Robustness under Traffic Surge}
To evaluate the utility of the algorithm under different service request pressures, we fix the mean task lifetime at $15$ time units and test performance across varying arrival rates. As illustrated in Fig. \ref{fig:hops_time_1}, the average hops $r_t$ for all algorithms exhibit an upward trend as the arrival rate increases. This phenomenon occurs because high traffic volume exacerbates network congestion. It will deplete continuous idle resources and force workflows to be deployed across dispersed nodes.

However, AgentVNE demonstrates superior resilience with the most gradual growth curve. Under high-load scenarios, AgentVNE maintains the weighted average hops at a significantly lower level. It is approximately 30\% of that yielded by the baseline algorithms, equivalent to a $\sim$70\% reduction in communication overhead.

Moreover, this performance advantage stems from two synergistic mechanisms:
\begin{itemize}
    \item RL-based Congestion Avoidance: The congestion penalty mechanism embedded in our RL reward function (Eq. \ref{eq:reward}) enables the PPO agent to proactively identify and avoid potential network bottlenecks. It guides placements toward topologically compact regions.
    \item LLM-driven Affinity Anchoring: The resource augmentation mechanism captures affinity dependencies and injects virtual resource bias into constraint-satisfying physical nodes. This artificially elevates the ranking of these nodes and their neighbors in the candidate priority list. Thus, it creates a ``gravitational pull'' that guides the workflow toward these anchor nodes.   Since specific agents are mandatorily bound to these affinity nodes. Then, we cluster the remaining virtual nodes in their vicinity to minimize the physical distance between agents. This further reduces the global communication hop count.
\end{itemize}


\subsubsection{Impact of Lifecycle and Anti-Fragmentation}
We further investigate the impact of task duration by fixing the arrival rate at 0.2 and varying the workflow mean lifetime (Fig. \ref{fig:hops_time_2}). As the task lifetime extends, the resource release cycle slows down and becomes unevenly distributed, leading to severe resource fragmentation.

The performance gap widens as the mean lifetime increases from 15 to 40. In long-duration scenarios, AgentVNE maintains the weighted average hops at approximately 44\% of the level exhibited by baseline algorithms. It demonstrates superior anti-fragmentation capability.

Meanwhile, this advantage stems from RL-driven long-term planning. The PPO agent learns a fragmentation-aware compact placement strategy. It proactively schedules smaller sub-tasks onto scattered resource fragments, thereby preserving large contiguous physical resource blocks for complex collaborative workflows. In contrast, traditional greedy strategies, driven by myopic decision-making, tend to occupy large contiguous resource blocks indiscriminately for current requests. It forces subsequent tasks to be deployed across dispersed, fragmented nodes with high communication costs.

\begin{CJK*}{UTF8}{gbsn}

\end{CJK*}

\subsection{Service Acceptance Ratio Analysis}
The service acceptance rate (SAR) is a crucial metric for evaluating the efficacy of the workflow placement strategy.
This strategy aims to successfully deploy MASs.
It also focuses on scheduling resources under constrained conditions.


\begin{figure*}[htbp]
    \centering
    \subfloat[Long-term Accepted Ratio\label{fig:vn}]{
        \includegraphics[width=0.31\textwidth]{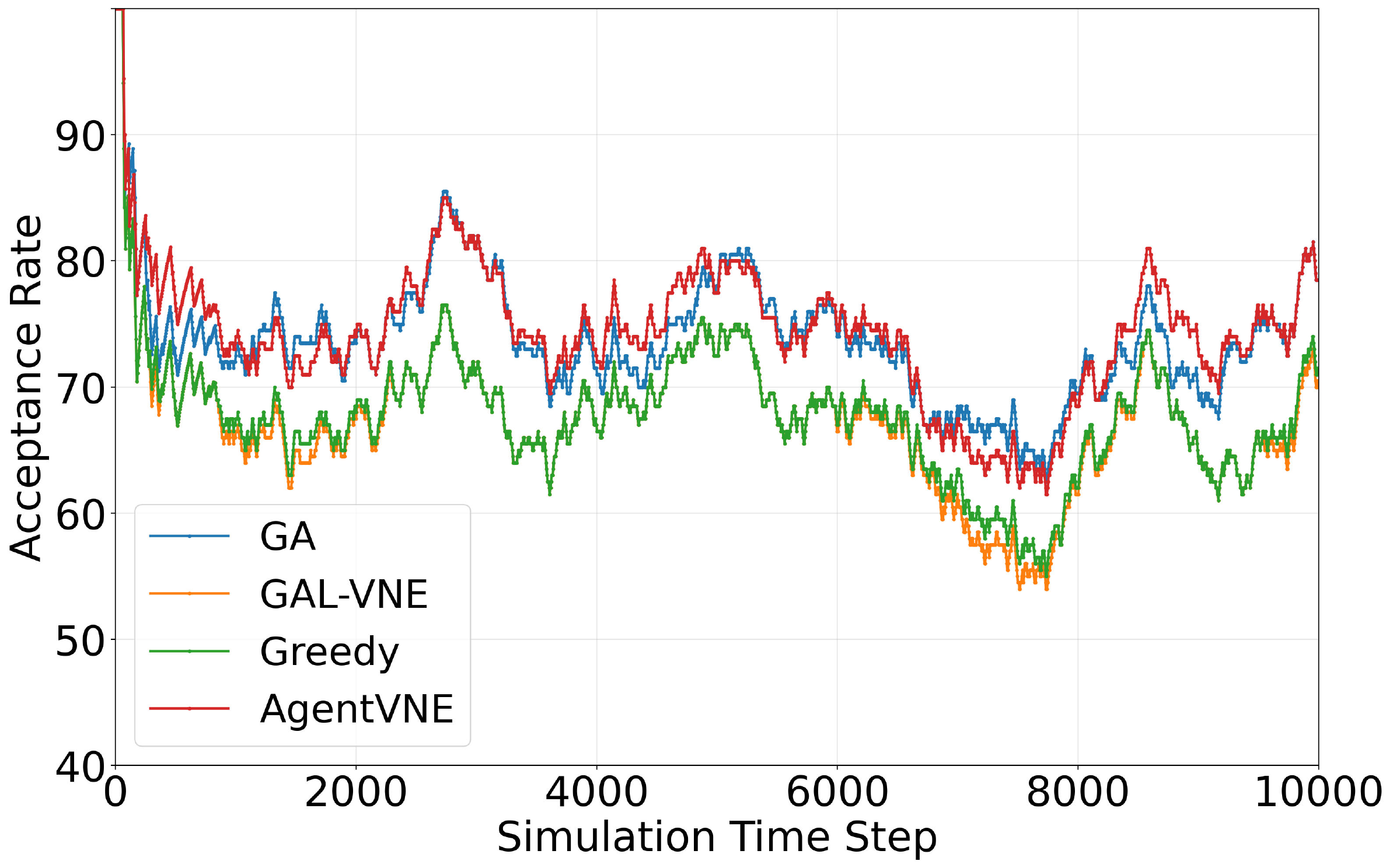}
        \label{fig:ac_time}
    }
    \subfloat[Arrival Rate\label{fig:pretrain_prob}]{
        \includegraphics[width=0.31\textwidth]{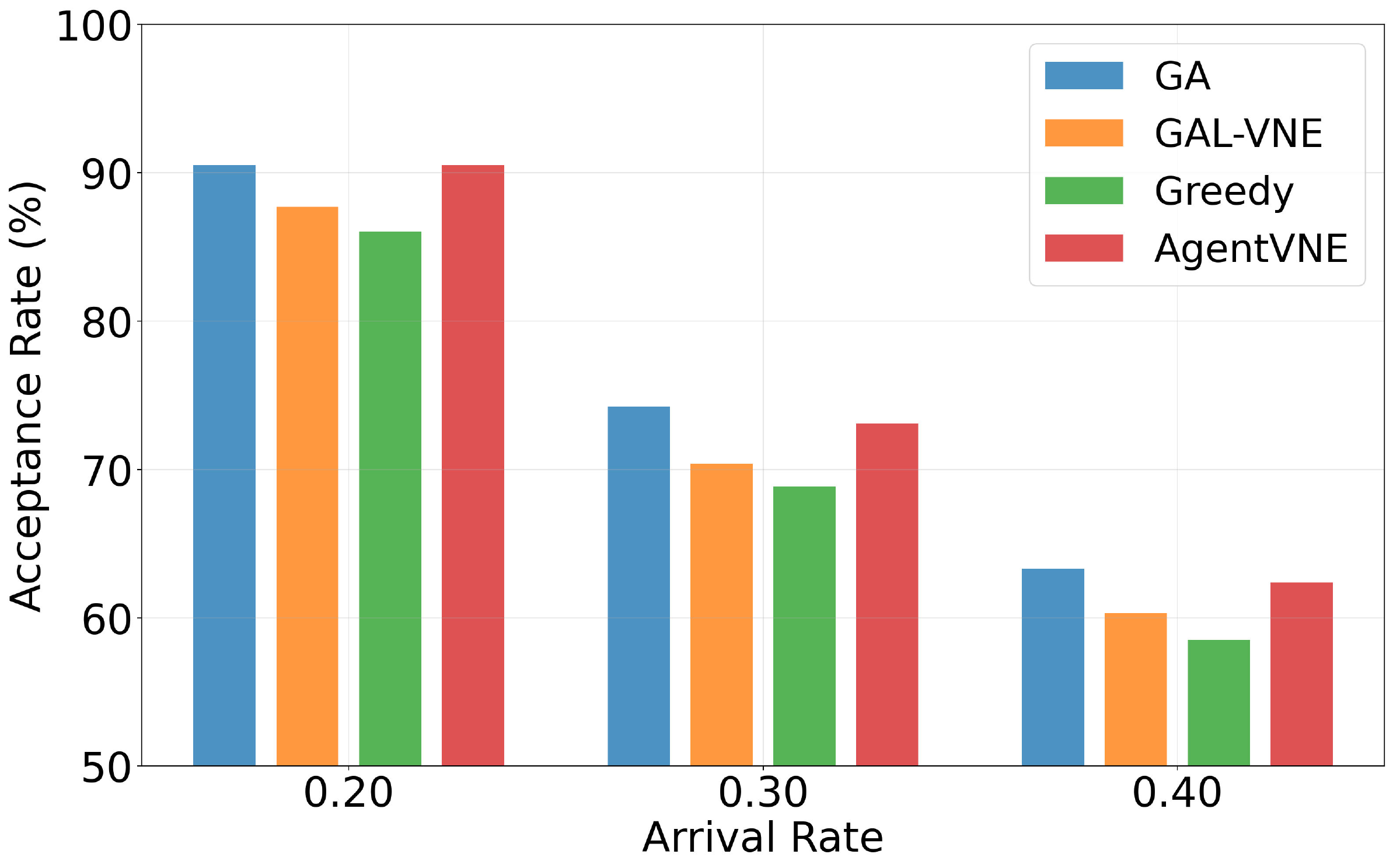}
        \label{fig:ac_ar}
    }
    \subfloat[Workflow Mean Lifetime\label{fig:finetuned_prob}]{
        \includegraphics[width=0.31\textwidth]{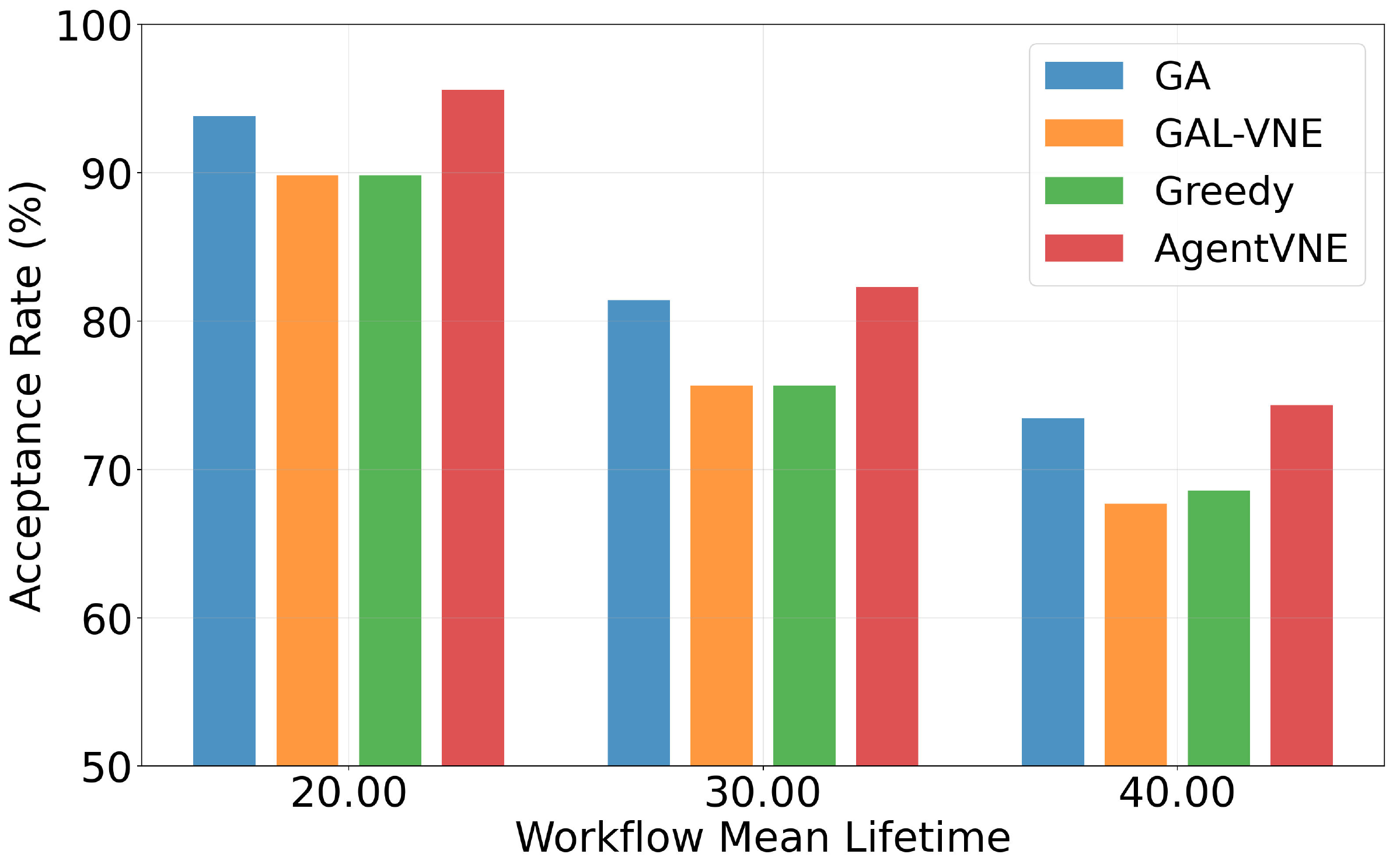}
        \label{fig:ac_lt}
    }
    \caption{Multi-agent workflow acceptance rates comparison, namely deployment success rates in different scenarios}
    \label{fig:a_c}
    \vspace{-0.5em}  
\end{figure*}



\subsubsection{Temporal Stability and Constraint Adaptation}
Fig. \ref{fig:ac_time} illustrates the variation of SAR over simulation time. AgentVNE consistently maintains the highest acceptance rate throughout the cycle, significantly outperforming GAL-VNE and traditional heuristics. 
AgentVNE's resource bias bias mechanism effectively pre-resolves these matching challenges. Furthermore, the PPO fine-tuning mechanism empowers the model to continuously adapt to dynamic traffic fluctuations, ensuring sustained high acceptance rates.
\subsubsection{Robustness under Traffic Surge}
To evaluate SAR performance under varying service request pressures, we fix the mean task lifetime at $30$ time units and test across different arrival rates, as shown in Fig. \ref{fig:ac_ar}. As the arrival rate increases from $0.2$ to $0.4$, network congestion intensifies, and idle resources become scarce, causing SAR to decline for all algorithms. However, AgentVNE exhibits the most gradual decrease, maintaining a SAR approximately $5\%-10\%$ higher than baselines at high load, with an arrival rate of $0.4$. This advantage stems from its ability to perceive multi-dimensional geometric similarity between virtual resource demands and physical supplies. It minimizes resource fragmentation through compact placement and delaying network saturation.

\subsubsection{Impact of Lifecycle and Anti-Fragmentation}
We further explore the influence of task duration on SAR by fixing the arrival rate at $0.25$ and varying the workflow mean lifetime, as shown in Fig. \ref{fig:ac_lt}. As the task lifetime extends from $20$ to $40$ time units, resource release cycles slow down, leading to severe resource fragmentation. Consequently, traditional heuristic algorithms, limited by short-sighted decision-making, suffer a sharp decline in SAR. In contrast, AgentVNE leverages the long-term reward mechanism of PPO fine-tuning to efficiently utilize fragmented resources, maintaining near-optimal SAR even for long-duration tasks.

Crucially, this resilience is underpinned by the NTN-enabled Multi-dimensional Alignment mechanism. The NTN explicitly evaluates the similarity between virtual and physical nodes across diverse resource dimensions. This capability enables the system to proactively schedule small-scale sub-tasks onto micro-fragments with matching resource profiles. By ensuring dimensional alignment, this mechanism effectively prevents single-dimension exhaustion. That is a scenario where a physical node becomes unusable due to the depletion of one resource type (e.g., CPU) despite having ample capacity in others. Thus, it can maximize the utilization of fragmented resources. Consequently, the scattered idle resources within the network are more fully exploited. This efficient resource consolidation directly translates into a superior SAR, empowering the edge infrastructure to accommodate a significantly larger volume of concurrent agentic tasks.


\subsection{Scalability Analysis}
We design scalability simulations to verify AgentVNE’s ability to optimize embedding quality via the resource diversity of large-scale networks. We also use these simulations to confirm its capacity to maintain efficient decision-making speed as network nodes and task loads grow.


\begin{table}[tbp]
  \centering
  \caption{Scalability Analysis: Impact of Network Size on Workflow Latency under Proportional Load}
  \label{tab:scalability}
  \renewcommand{\arraystretch}{1.2}
  \begin{tabular}{ccc} 
    \hline
    \textbf{Substrate Network Size} & \textbf{Arrival Rate} & \textbf{Average Hops} \\
    \hline
    20  & 0.2 & 4.6 \\
    50  & 0.5 & 4.1 \\
    100 & 1.0 & 3.5 \\
    150 & 1.5 & 2.9 \\
    200 & 2.0 & 2.2 \\
    \hline
  \end{tabular}
\end{table}

\subsubsection{Impact of Network Size on Communication Overhead}
To evaluate the scalability of AgentVNE, we conduct a stress test. It sets the number of substrate nodes ranging from $20$ to $200$, adopting a proportional load scaling strategy. The task arrival rate increases linearly with the number of nodes. The experimental results, as shown in Table \ref{tab:scalability}, indicate that as the network scale expands, the weighted average hops of workflows decrease from $4.6$ to $2.2$. This demonstrates that AgentVNE effectively leverages the denser solution space in large-scale networks. Then, it can identify node clusters within local neighborhoods that satisfy both topological and affinity constraints, thereby achieving a more compact embedding.

\subsubsection{Solving Efficiency}
\begin{figure}[tbp]
    \centering
    \includegraphics[width=0.88\linewidth]{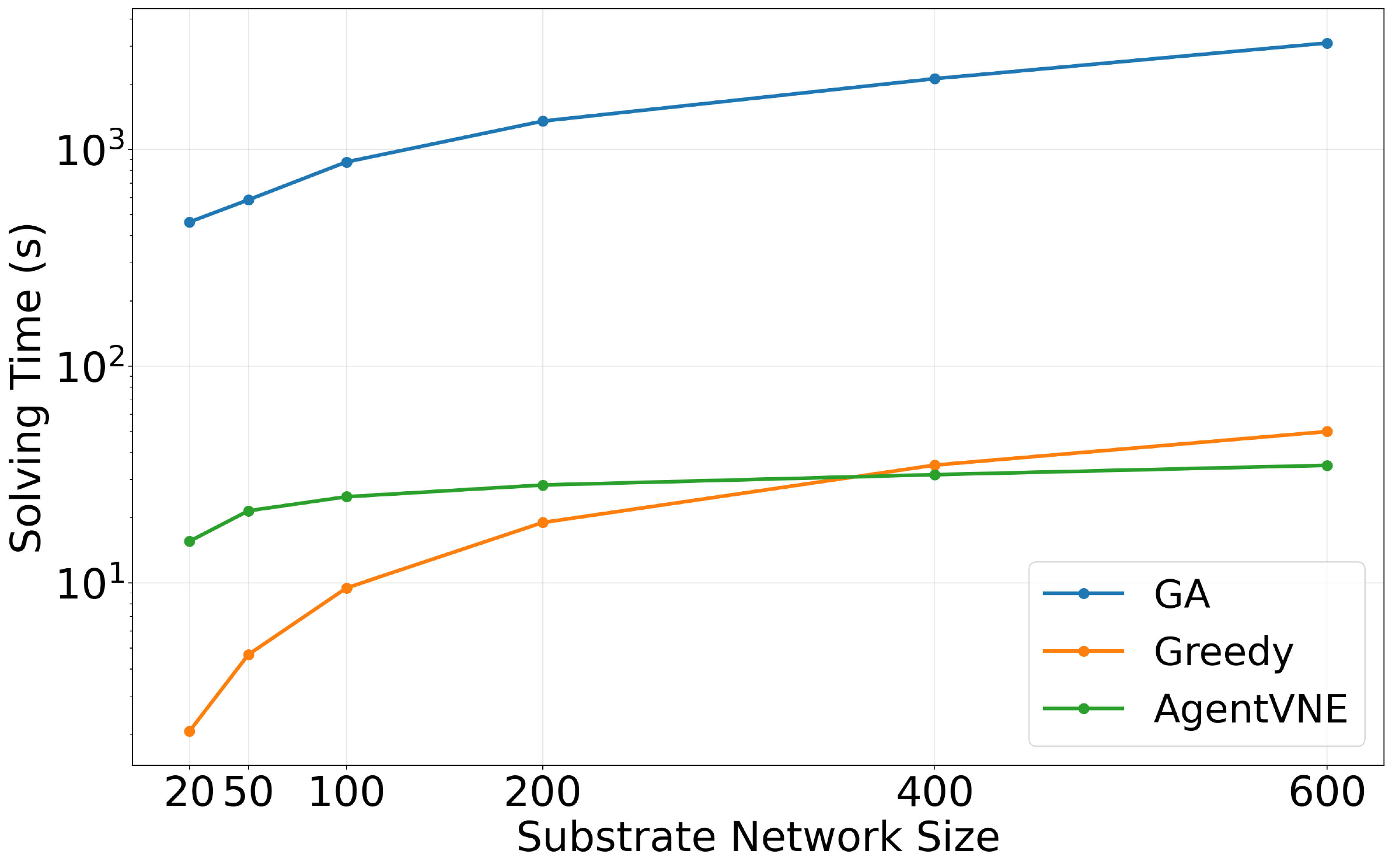}
    \caption{Variation of solving time over substrate network size .}
    \label{fig:solving_time}
\end{figure}

Fig. \ref{fig:solving_time} illustrates the average solving time for processing 1,000 tasks across different algorithms. Although meta-heuristic algorithms can achieve respectable placement quality through iterative search in complex dynamic environments, their solving time grows quadratically with network size, often exceeding tens of seconds. For real-time agentic workflows, this high computational latency may exceed the communication latency saved by the optimization. In contrast, AgentVNE benefits from the parallel inference capabilities of neural networks. Despite a fixed overhead at small scales, its solving time increases marginally with network size. In large-scale networks, for multi-agent workflows that do not require LLM to capture placement constraints, AgentVNE generates near-optimal solutions within seconds. It can achieve an optimal balance between solution quality and decision latency, thus demonstrating its practical value in latency-sensitive edge applications.

\subsection{Ablation Evaluation}

To validate the effectiveness of the core modules within the AgentVNE dual-layer architecture, we design ablation experiments to separately evaluate the impact of the LLM-assisted constraint perception module and the Reinforcement Learning (RL)-based dynamic fine-tuning mechanism on system performance.

\subsubsection{Impact of LLM Module} 
We construct a variant denoted as $\text{AgentVNE}_{w/oLLM}$, which removes the semantic analysis layer and disables the generation of the resource bias.
As shown in Table \ref{tab:ablation}, after removing the LLM module, although the algorithm maintains a high SAR, its workflow weighted average hops $r_t$ increase significantly. For example, at an arrival rate of $0.2$, the average hops of $\text{AgentVNE}_{w/oLLM}$ reach $8.4$, which is approximately $1.8$ times that of the complete AgentVNE with $4.6$ hops.
This indicates that without the guidance of the semantic layer, the model struggles to identify implicit hardware affinity constraints. Thus, it will result in agents being dispersed across nodes with abundant physical resources but large topological distances. Conversely, with the introduction of the LLM's physical resource bias injection mechanism, the system can effectively identify constraint anchors and shift the center of sampling probability toward the neighborhood of restricted nodes. Then, it can reduce communication overhead by approximately $30\%$ to $40\%$.

\subsubsection{Impact of RL Fine-tuning} 
To evaluate the necessity of PPO fine-tuning, we denote the pre-trained model without RL updates as $\text{AgentVNE}_{w/oRL}$.
Experimental data reveal that $\text{AgentVNE}_{w/oRL}$ outperforms $\text{AgentVNE}_{w/oLLM}$ in hop performance due to the presence of LLM bias injection. However, its SAR drops noticeably when the task arrival rate is high, falling from $67\%$ to $63\%$.
This suggests that relying solely on a pre-trained model is insufficient to cope with highly dynamic task flows and resource fragmentation. The AgentVNE tailored via PPO fine-tuning can learn shape alignment strategies in the multi-dimensional resource space through continuous interaction with the environment. Thus, it can enhance the system's service carrying capacity while ensuring low latency.

\begin{table}[tbp]
  \centering
  \caption{Ablation Analysis: Impact of the LLM-Assisted Module on Workflow Latency and Service Acceptance Rate}
  \begin{tabular}{lccc}
    \hline
    \textbf{Solution} & \textbf{Arrival Rate} & \textbf{Average Hops} & \textbf{AC Ratio} \\
    \hline
    $\text{AgentVNE}_{w/oLLM}$ & 0.2 & 8.4 & 97\% \\
    $\text{AgentVNE}_{w/oLLM}$ & 0.4 & 10.8 & 66\% \\
    $\text{AgentVNE}_{w/oLLM}$ & 0.6 & 15.2 & 46\% \\
    $\text{AgentVNE}_{w/oRL}$ & 0.2 & 6.2 & 95\% \\
    $\text{AgentVNE}_{w/oRL}$ & 0.4 & 9.8 & 63\% \\
    $\text{AgentVNE}_{w/oRL}$ & 0.6 & 14.2 & 43\% \\
    \rowcolor{green!15} \text{AgentVNE} & 0.2 & 4.6 & 97\% \\
    \rowcolor{green!15} \text{AgentVNE} & 0.4 & 7.2 & 67\% \\
    \rowcolor{green!15} \text{AgentVNE} & 0.6 & 10.4 & 49\% \\
    \hline
  \end{tabular}
  \label{tab:ablation}
\end{table}

\begin{figure}[tbp]
    \centering
    \includegraphics[width=0.68\linewidth]{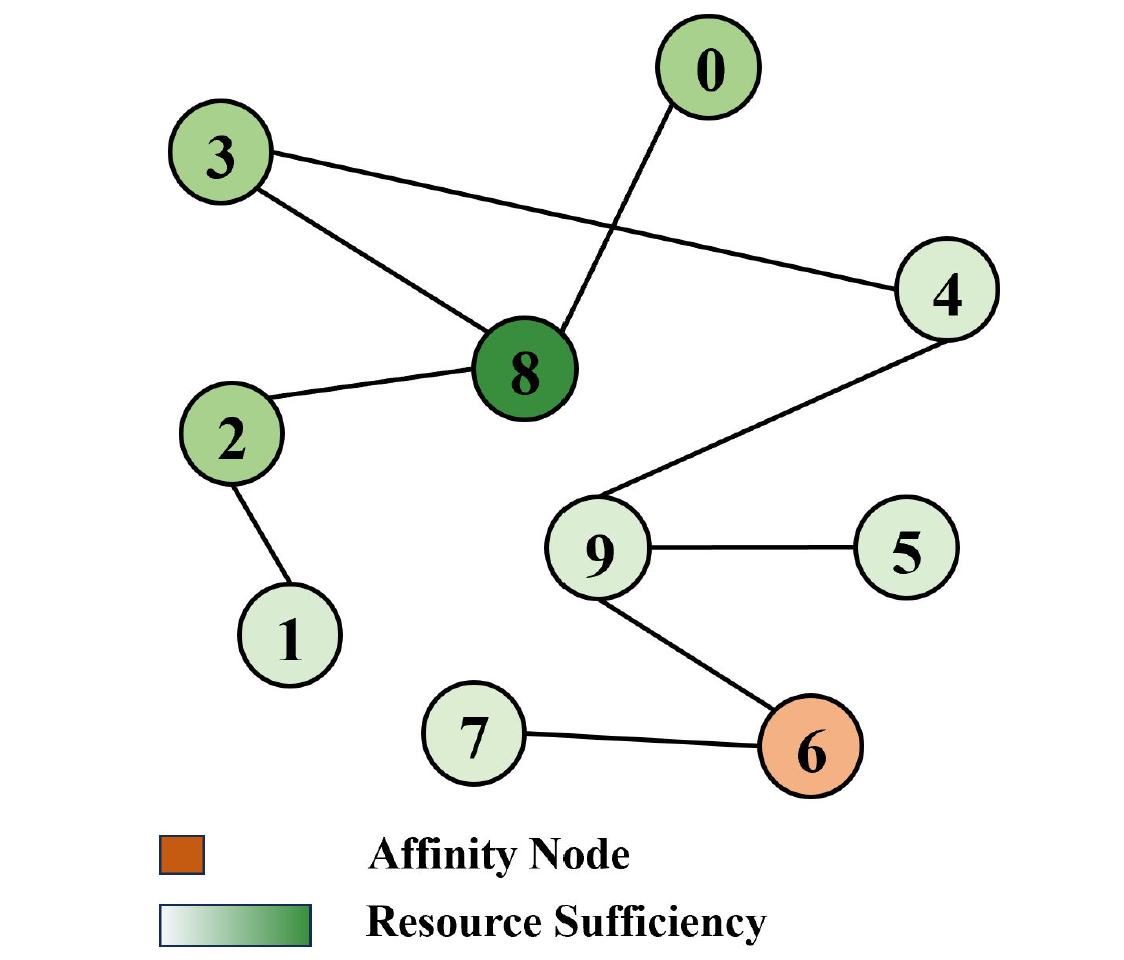}
    \caption{Substrate nodes topology visualization. The darker the color, the more abundant the resources.}
    \label{fig:vn_topo}
\end{figure}

\begin{figure*}[htbp]
    \centering
    \subfloat[\small Node mapping probability distribution during pre-training phase\label{fig:pretrain_prob}]{
        \includegraphics[width=0.42\textwidth]{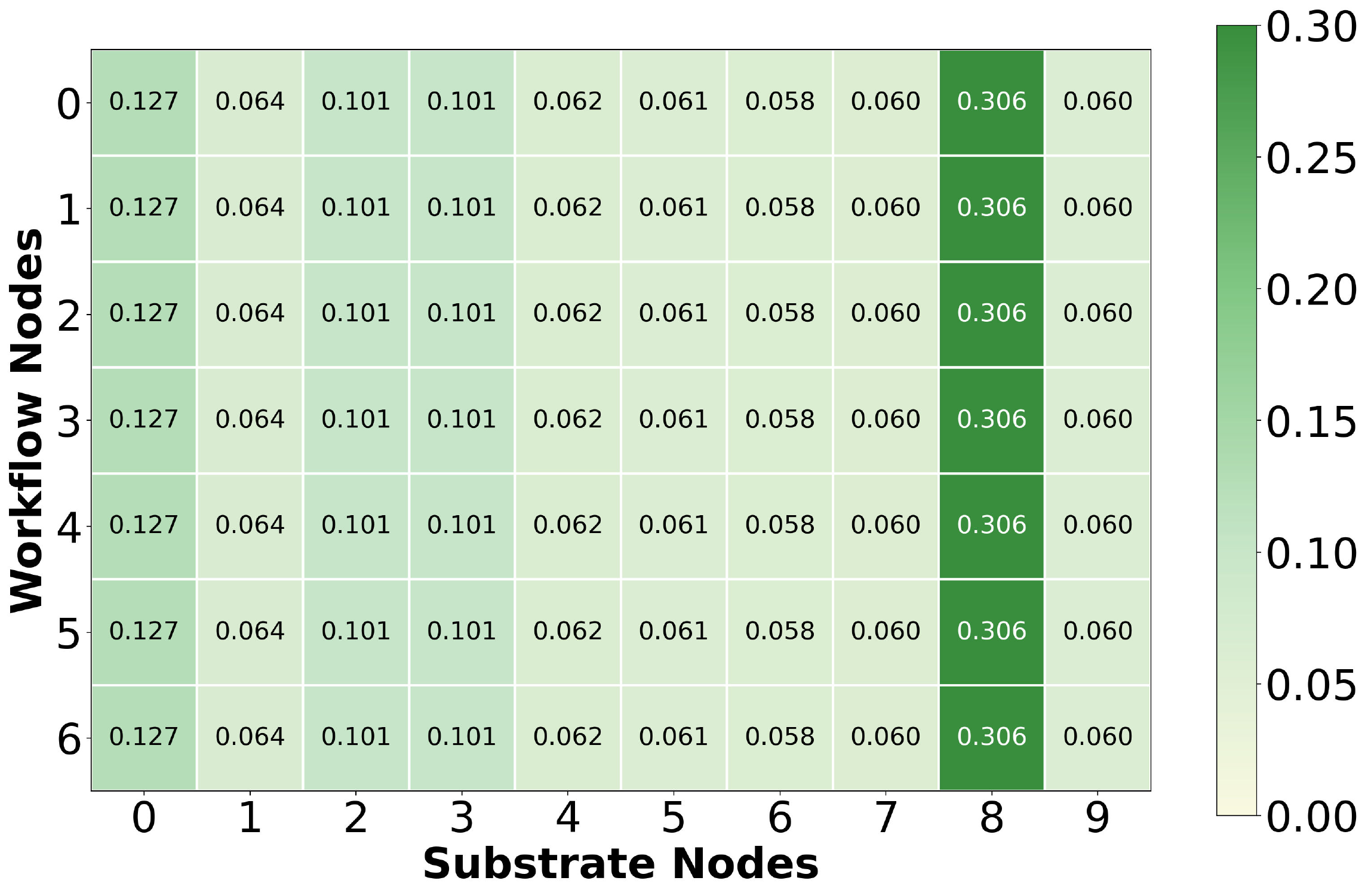}
    }
    \hspace{0.4em} 
    \subfloat[\small Node mapping probability distribution after finetuning\label{fig:finetuned_prob}]{
        \includegraphics[width=0.42\textwidth]{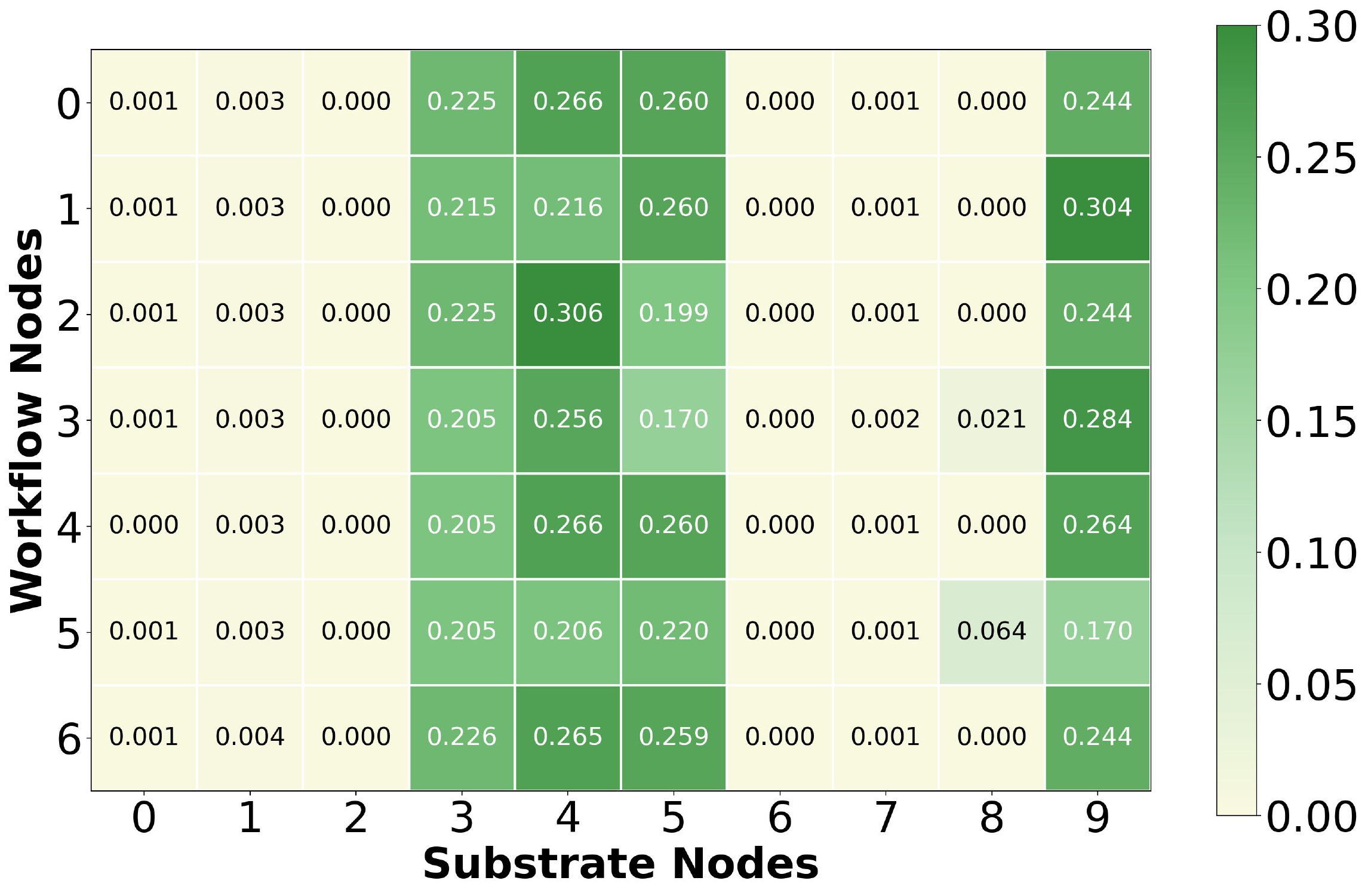}
    }
    \caption{Comparison of node mapping probability matrices. 
    }
    \label{fig:prob_comparison}
    \vspace{-0.5em}  
\end{figure*}

Then, to visually demonstrate the decision logic of AgentVNE, we visualize a substrate nodes topo in Fig. \ref{fig:vn_topo} and the node mapping probability matrices output by the model in Fig. \ref{fig:prob_comparison}. In the setup, the workflow contains an agent that must be placed on substrate node $6$.

\begin{itemize}
    \item Pre-training phase: As shown in Fig. \ref{fig:pretrain_prob}, the model tends to mimic a greedy strategy, assigning extremely high sampling probability to Node $8$, which possesses the most abundant residual resources. However, it ignores the fact that Node $8$ is topologically far from the anchor Node $6$, leading to potential long-tail communication latency.
    \item Fine-tuning phase: As shown in Fig. \ref{fig:finetuned_prob}, AgentVNE rectifies the probability distribution. Despite node $8$'s high capacity, its probability is suppressed due to excessive topological distance. In contrast, nodes $3$, $4$, and $5$, which are immediate neighbors of the anchor node $6$, receive significantly improved sampling probabilities. This proves that AgentVNE has successfully learned to prioritize selecting topologically affine neighbor nodes while satisfying resource constraints. Thus, it can achieve a more compact topological embedding.
\end{itemize}

\section{Related Work}\label{sec:relatedwork}

\begin{table*}[htbp]
\caption{Comparison of AgentVNE with Existing Related Works}
\label{tab:related_work_comparison}
\centering
\resizebox{\textwidth}{!}{%
\begin{tabular}{llllcc}
\toprule
\textbf{Reference} & \textbf{Main Scenario} & \textbf{Service Object} & \textbf{Methodology} & \textbf{Semantic Awareness} & \textbf{Dynamic Topology} \\
\midrule
Fu et al. \cite{fu2019dynamic} & NFV-IoT & SFC / VNF & DRL (DQN) & No & Low \\
Lin et al. \cite{lin2021optimizing} & Mobile Edge Intel. & AI Service Program & MINLP / ADMM & No & Low \\
Xia et al. \cite{xia2024privacy} & MEC & General Task & DRL (DDPG) & No & Low \\
Liu \& Zhang \cite{liu2024service} & Cloud/NFV & SFC & DRL (DDPG) & No & Medium \\
Wang et al. \cite{wang2025adaptive} & Edge Intelligence & AI Agent & Ant Colony + LLM & Limited & Medium \\
Liu et al. \cite{hera2025hybrid} & Hybrid Edge-Cloud & Agent Subtasks & Heuristic Scheduler & No & High \\
\midrule
\textbf{AgentVNE (Ours)} & \textbf{Edge Agentic AI} & \textbf{Multi-Agent Workflow} & \textbf{LLM + DRL (PPO)} & \textbf{High} & \textbf{High} \\
\bottomrule
\end{tabular}%
}
\end{table*}

\begin{CJK*}{UTF8}{gbsn}

VNE serves as a core technology in network virtualization. Early research predominantly focused on resource optimization in static environments, utilizing heuristic or meta-heuristic algorithms to address node and link mapping problems~\cite{zhang2024qos, yao2018novel}. With the proliferation of Network Function Virtualization (NFV), the research focus has gradually shifted towards the dynamic embedding of service function chains (SFC). For instance, Liu and Zhang~\cite{liu2024service} proposed a DRL-based SFC embedding framework that employs BERT to encode network states for handling traffic fluctuations, while Fu et al.~\cite{fu2019dynamic} designed a dependency-aware SFC mapping strategy specifically for IoT scenarios.

Moreover, to deploy AI services at the resource-constrained edge, extensive exploration has been conducted in model compression and collaborative inference. Chen et al.~\cite{chen2024spaceedge} proposed an efficient offloading paradigm for computation-intensive tasks in space-air-ground integrated networks; Yin et al.~\cite{yin2025elastic} explored elastic scaling strategies for LLM services on edge devices, analyzing the fluctuation characteristics of inference workloads. Additionally, Kim et al.~\cite{hera2025hybrid} introduced a hybrid edge-cloud resource allocation mechanism to optimize inference cost and accuracy by combining SLMs with LLMs. However, most of these works treat AI tasks as independent computational units or unidirectional data flow pipelines~\cite{li2024incentive}, neglecting the complex semantic interactions between agents in Agentic Systems. The topology of such systems is often a generative graph topology~\cite{zhang2025evoflow}, dynamically generated by the planner at runtime based on inference logic. This topological deformation, which evolves with the CoT, makes algorithms based on predefined SFCs difficult to adapt directly.

Furthermore, with the rise of the IoA concept, resource scheduling for MASs has become a frontier hotspot. Early work demonstrated the feasibility of using LLMs to simulate multi-agent interactions and emergent behaviors~\cite{park2023generative}. Subsequently, frameworks such as AgentVerse~\cite{chen2024agentverse} and AutoGen~\cite{wu2024autogen} provided the necessary infrastructure for agent collaboration. Chen et al.~\cite{chen2025internet} further facilitated collaboration among heterogeneous LLM-based agents by introducing agent integration protocols, an instant-messaging-like architecture, and dynamic team-building mechanisms. Regarding resource allocation, Lin et al.~\cite{lin2021optimizing} investigated service placement in mobile edge intelligence but primarily focused on general energy optimization. Although Fan et al.~\cite{fan2025dynamic} proposed a dynamic topology resource allocation scheme for distributed training tasks, the parameter synchronization mode of distributed training differs fundamentally from the asynchronous interaction mode of agent inference. This makes it unsuitable for direct application in agent-based scenarios. Xia et al.~\cite{xia2024privacy} proposed a DRL-based privacy-preserving offloading framework, and Luo et al.~\cite{luo2025toward} discussed trust architectures in multi-LLM collaboration. Wang et al.~\cite{wang2025adaptive} incorporated ant colony algorithms for agent placement but only considered single-entity capacity matching. The proposed AgentVNE aims to bridge this gap by leveraging GNNs to capture high-order topological similarities, thereby achieving end-to-end optimization for agentic workflows.

\end{CJK*}

\section{Conclusion}\label{sec:conclusion}
In this paper, we have proposed AgentVNE, a cloud-edge collaborative framework designed to orchestrate MAS within the IoA. Addressing the limitations of traditional embedding algorithms, AgentVNE introduces a dual-layer architecture that synergizes LLMs with GNNs. Specifically, we utilize an LLM-driven resource bias mechanism to resolve hard physical constraints and a similarity-aware GNN optimized via pre-training and PPO fine-tuning to handle topological dynamism. Extensive simulations have demonstrated that AgentVNE significantly outperforms existing baselines, reducing communication latency to less than 40\% of that yielded by state-of-the-art methods while improving the service acceptance rate by approximately 5\%--10\% under high-load scenarios. This work provides a foundational solution for the semantic-aware deployment of agentic AI. Future work will focus on cross-domain agent migration.

\bibliographystyle{IEEEtran} 
\bibliography{IEEEabrv,reference_2}

\end{document}